\documentclass[12pt]{article}%

\renewcommand\footnotemark{}

\usepackage{amssymb}
\usepackage{amsfonts}
\usepackage{amsmath}
\usepackage[nohead]{geometry}
\usepackage[singlespacing]{setspace}
\usepackage[bottom]{footmisc}
\usepackage{indentfirst}
\usepackage{endnotes}
\usepackage{graphicx}%
\usepackage{rotating}
\usepackage{verbatim}
\usepackage{setspace}
\usepackage{multirow}
\usepackage{latexsym}
\usepackage{bbm}
\usepackage{enumitem}
\usepackage{hhline}

\usepackage{graphicx}
\usepackage{caption}
\usepackage{subcaption}

\usepackage{amsthm}
\usepackage{makeidx}
\usepackage{fancyhdr}
\usepackage{type1cm}

\usepackage[authoryear]{natbib}

\newcommand{\tr}{\mathrm{tr}}
\newcommand{\diag}{\mathrm{diag}}
\newcommand{\argmin}{\mathrm{argmin}}

\def \Var {\mbox{Var}}
\def \Cov {\mbox{Cov}}
\def \diag {\mbox{diag}}
\def \rank {\mbox{rank}}
\def \sgn {\mbox{sgn}}

\newcommand{\beq}{\begin{equation}}
\newcommand{\eeq}{\end{equation}}
\newcommand{\beqn}{\begin{eqnarray}}
\newcommand{\eeqn}{\end{eqnarray}}
\newcommand{\beqnn}{\begin{eqnarray*}}
\newcommand{\eeqnn}{\end{eqnarray*}}

\numberwithin{equation}{section}
\theoremstyle{plain}
\newtheorem{thm}{Theorem}[section]

\newtheorem{lem}{Lemma}[section]
\newtheorem{cor}{Corollary}[section]

\newtheorem{assmp}{Assumption}[section]
\theoremstyle{definition}

\makeatletter
\def\@biblabel#1{\hspace*{-\labelsep}}
\makeatother
\begin{document}

\title{Robust Covariance Estimation for Approximate Factor Models}
\author{Jianqing Fan$^*$
\thanks{Address: Department of ORFE, Sherrerd Hall, Princeton University, Princeton, NJ 08544, USA, e-mail: \textit{jqfan@princeton.edu}, \textit{yiqiaoz@princeton.edu},
\textit{weichenw@princeton.edu}. The research was partially supported by NSF grants
DMS-1206464 and DMS-1406266 and NIH grants R01-GM072611-11
and NIH R01GM100474-04.}, Weichen Wang$^*$ and Yiqiao Zhong$^*$
\medskip\\{\normalsize $^*$Department of Operations Research and Financial Engineering,  Princeton University}
}

\date{}

\maketitle

\sloppy

\onehalfspacing

\begin{abstract}
In this paper, we study robust covariance estimation under the approximate factor model with observed factors. We propose a novel framework to first estimate the initial joint covariance matrix of the observed data and the factors, and then use it to recover the covariance matrix of the observed data. We prove that once the initial matrix estimator is good enough to maintain the element-wise optimal rate, the whole procedure will generate an estimated covariance with desired properties. For data with only bounded fourth moments, we propose to use Huber loss minimization to give the initial joint covariance estimation. This approach is applicable to a much wider range of distributions, including sub-Gaussian and elliptical distributions.  We also present an asymptotic result for Huber's M-estimator with a diverging parameter. The conclusions are demonstrated by extensive simulations and real data analysis.
\end{abstract}

\textbf{Keywords:} Robust covariance matrix, Approximate factor model, M-estimator.

\pagebreak%
\doublespacing

\onehalfspacing

\section{Introduction}\label{sec::intro}

The problem of estimating a covariance matrix and its inverse has been fundamental in many areas of statistics, including principal component analysis (PCA), linear discriminative analysis for classification, and undirected graphical models, just to name a few. The intense research in high dimensional statistics has contributed a stream of papers related to covariance matrix estimation, including sparse principal component analysis \citep{JohLu09, AmiWai09, VuLei12, BirJohNadPau13, BerRig13, Ma13, CaiMaWu13}, sparse covariance estimation \citep{BicLev08, CaiLiu11, CaiZhaZho10, LamFan09, RavWaiRasYu11} and factor model analysis \citep{StoWat02, Bai03, FanFanLv08, FanLiaMin13, FanLiaWan14, Ona12}. A strong interest in precision matrix estimation (undirected graphical model) has also emerged in the statistics community following the pioneering works in \cite{MeiBuh06} and \cite{FriHasTib08}. In the application aspect, many areas such as portfolio allocation \citep{FanFanLv08}, have benefited from this continuing research.

In the high dimensional setting, the number of variables $p$ is comparable or greater than the sample size $n$. This dimensionality poses a challenge to the estimation of covariance matrices. It has been shown in \cite{JohLu09} that the empirical covariance matrix behaves poorly, and sparsity of leading eigenvectors is assumed to circumvent this issue. Following this work, a flourishing literature on sparse PCA has developed in-depth analysis and refined algorithms; see \cite{VuLei12, BerRig13, Ma13}. Taking a different route, \cite{BicLev08} advocated thresholding as a regularization approach to estimate a sparse matrix, in the sense that most entries of the matrix are close to zero.

Another challenge in high-dimensional statistics is that the measurements can not have light tails, as large scale data are often obtained by using (bio)imaging technologies that have a limited precision.  Moreover, it is well known that financial returns exhibit heavy tails.  These invalidate the fundamental assumptions in high-dimensional statistics that data have sub-Gaussian or sub-exponential tails, popularly imposed in most of the aforementioned papers. Significant relax of the above assumption requires some new ideas and forms the subject of this paper.

Recently, motivated by Fama-French model \citep{FamFre93} from financial econometrics, \cite{FanFanLv08} and \cite{FanLiaMin13} considered the covariance structure of the \textit{static approximate factor model}, which models the covariance matrix by a low-rank signal matrix and a sparse noise matrix. The same model will also be the focus of this paper. The model assumes existence of  several low-dimensional factors that drives a large panel data $\{y_{it}\}_{i\le p, t\le n}$, that is
\begin{equation}\label{eq1.1}
y_{it} = b_i^T f_t + u_{it}\,, \quad \quad i \le p, \; t \le n,
\end{equation}
where $f_t$'s are the common factors and $b_i$'s are their corresponding factor loadings. The noises $u_{it}$'s, known as the idiosyncratic component, are uncorrelated with the factors $f_t \in \mathbf{R}^r$. Here $r$ is relatively small compared with $p$ and $n$. We will treat $r$ as fixed independent of $p$ and $n$ throughout this paper. When the factors are known, this model subsumes the well-known CAPM model \citep{Sha64, Lin65} and Fama-French model \citep{FamFre93}. When $f_t$ is unobserved, the model tries to recover the underlying factors for the movements of the whole panel data. Here the approximate factor model means that the covariance $\Sigma_u$ of $u_t= (u_{1t}, \dots, u_{pt})$ is sparse, including the strict factor model in which $\Sigma_u$ is diagonal as a special case. In addition, ``static'' is on the contrary of the dynamic model which takes into account the time lag and allows more general infinite dimensional representations \citep{ForHalLipRei00, ForLip01}.

The covariance matrix of the outcome $y_t = (y_{1t}, \dots, y_{pt})'$ from model \eqref{eq1.1} can be written as
\begin{equation} \label{eq1.2}
\Sigma = B\Sigma_f B^T + \Sigma_u\,,
\end{equation}
where $B_{p \times r}$ consisting of $b_i'$ in each row is the loading matrix, $\Sigma_f$ is the covariance of $f_t$ and $\Sigma_u$ is the sparse covariance matrix for $u_t$. Here we assume the process of $(f_t,u_t)$ is stationary so that $\Sigma_f, \Sigma_u$ do not depend on time. When factors are unknown, \cite{FanLiaMin13} proposed applying PCA to obtain an estimate of the low rank part and sparse part $\Sigma$. The crucial assumption is that the factors are \textit{pervasive}, meaning that the factors have non-negligible effects on a large amount of dimensions of the outcomes. 
\cite{FanWan15} gives more explanation from random matrix theories and aims to relax the pervasiveness assumption in applications such as risk management and estimation of the false discovery proportion. See \cite{Ona12} for more discussions on strong and semi-strong factors.

In this paper, we consider estimating $\Sigma$ simply with known factors. The main focus of the paper is on robustness instead of factor recovery. Under exponential tails of the factors and noises, \cite{FanLiaMin11} proposed the idea of performing thresholding on the estimate of $\Sigma_u$, obtained from the sample covariance of the residuals of multiple regression \eqref{eq1.1}.
The legitimacy of this approach hinges on the assumption that the tails of the factor and error distributions are exponential decay, which is likely to be violated in practice, especially in the financial applications. Thus, the need to extend the applicability of this approach beyond well-behaved noise has driven further research such as \cite{FanLiuWan15}, in which they assume that $y_t$ has an elliptical distribution \citep{FanKotNg90}. 

This paper studies the model (\ref{eq1.1}) under a much more relaxed condition, that the random variables $f_t$ and $u_{it}$ only have finite fourth moments.
The main observation that motivates our method is that, the joint covariance matrix of $(y_t^T, f_t^T)^T$ supplies sufficient information to estimate $B\Sigma_f B^T$ and $\Sigma_u$. To estimate the joint covariance matrix in a robust way, the classical idea that dates back to \cite{Hub64} proves to be vital and effective. The novelty here is that we let the parameter diverges in order to control the bias in high-dimensional applications. The Huber loss function with a diverging parameter, together with other similar functions, has been shown to produce concentration bounds for M-estimators, when the random variables have fat tails; see for example \cite{Cat12} and \cite{FanLiWan16}. This point will be clarified in Sections \ref{sec::est} and \ref{sec::theory}. The M-estimators considered here have additional merits in asymptotic analysis, which is studied in Section \ref{sec::asmp}.

This paper 
can be placed in the broader context of low rank plus sparse representation. In the past few years, robust principal component analysis has received much attention among statisticians, applied mathematicians and computer scientists. Their focus is on identifying the low rank component and sparse component from a corrupted matrix \citep{Cha11, CanLiMaWri11, XuCarSan10}. However, the matrices considered therein do not come from random samples, and as a result, neither estimation nor inference are involved. \cite{AgaNegWai12} does consider the noisy decomposition, but still, it focuses more on identifying and separating the low rank part and sparse part. In spite of connections with the robust PCA literature, such as the incoherence condition (see Section \ref{sec::est}), this paper and its predecessors are more engaged in disentangling ``true signal'' from noise, in order to improve estimation of covariance matrices. In this respect, they bear more similarity with the literature of covariance matrix estimation.

We make a few notational definitions before presenting the main results. For a general matrix $M$, the max norm of $M$, or the entry-wise maximum absolute value, is denoted as $\|M\|_{\infty} = \max_{ij}|M_{ij}|$. The operator norm of $M$ is $\| M \| = \lambda_{\text{max}}^{1/2}(M^TM)$ whereas the Frobenius norm is $\|M\|_F = \sqrt{\sum_{ij}M_{ij}^2}$. If $M$ is furthermore symmetric, we denote $\lambda_{j}(M)$ as the $j^{th}$ largest eigenvalue, $\lambda_{\text{max}}(M)$ as the largest one, and $\lambda_{\text{min}}(M)$ as the smallest one. In the paper, $C$ is a generic constant that may differ from line to line in the assumptions and also derivation of our theories.

The paper is organized as follows. In Section \ref{sec::est}, we present the procedure for robust covariance estimation when only finite fourth moment is assumed for both factors and noises without specific distribution family assumption. The theoretical justification will be provided in Section \ref{sec::theory}. Simulations will be carried out in Section \ref{sec::sim} to demonstrate the effectiveness of the proposed procedure. We also conduct real data analysis on portfolio risk of S\&P stocks via Fama-French model in Section \ref{sec::data}. Technical proofs will be delayed to the appendix.

\section{Robust covariance estimation}\label{sec::est}
Consider the factor model (\ref{eq1.1}) again with observed factors. It can be written in the vector form as
\begin{equation} \label{eq2.1}
y_t = B f_t + u_t \,,
\end{equation}
where $y_t = (y_{1t},\ldots, y_{pt})^T$, $f_t \in \mathbf{R}^r$ are the factors for $t = 1, \dots, T$, $B = (b_1, \dots, b_p)^T$ is the unknown loading matrix and $u_t = (u_{1t},\ldots, u_{pt})^T$ is uncorrelated with the factors. 
We assume that $(u_t^T,f_t^T)$ have zero mean and independent for $t = 1, \ldots, T$. A motivating example from economic and financial studies is the classical Fama-French model, where $y_{it}$'s represent excess returns of stocks in the market and $f_t$'s are interpreted as common factors driving the market. It is more natural to allow for weak temporal dependence such as $\alpha$-mixing as in the work of \cite{FanLiaWan14}. Though possible, we assume independence in this paper for the sake of simplicity of analysis.

\subsection{Assumptions} \label{sec2.1}

We now state the main assumptions of the model. Let $\Sigma_f$ be the covariance of $f_t$, and $\Sigma_u$ the covariance of $u_t$. A covariance decomposition shows that $\Sigma$, the covariance of $y_t$, comprises two parts,
\begin{equation}\label{eq2.2}
\Sigma = B \Sigma_f B^T + \Sigma_u\,.
\end{equation}
We assume that $\Sigma_u$ is sparse and the sparsity level is measured through
\begin{equation}
m_q = \max_{i \le p} \sum_{j \le p} (\Sigma_u)_{ij}^q, \quad \quad \text{for some } q \in [0,1].
\end{equation}
If $q=0$, $m$ is defined to be $\max_{i\le p}\sum_{j\le p} \mathbbm{1}((\Sigma_u)_{ij} \neq 0)$, i.e.\ the exact sparsity. An intuitive justification of the sparsity measurement stems from modeling of the covariance structure: after taking out the common factors, the rest only has weak cross-sectional dependence.
In addition, we assume that $\| \Sigma_u \|$, as well as $\|\Sigma_f\|$, is bounded away from $0$ and $\infty$. In the case of degenerate $\Sigma_f$, we can always consider rescaling the factors and reduce the number of observed factors to meet the requirement of non-vanishing minimum eigenvalue of $\Sigma_f$. This leads to our first assumption.
\begin{assmp}\label{assmp::Sigu}
There exists a constant $C>0$ such that $C^{-1} \le \|\Sigma_u \| \le C$ and $C^{-1} \le \| \Sigma_f \| \le C$, where $\Sigma_f$ is a $r\times r$ matrix with $r$ being a constant.
\end{assmp}

Furthermore, it is observed by \cite{StoWat02} that the factors are \textit{pervasive} in the sense that the low rank part of (\ref{eq2.2}) is the dominant component of $\Sigma$; more specifically, the top $r$ eigenvalues grow linearly as $p$. This motivates the following assumption:
\begin{assmp}\label{assmp::pervasive}
(i) There exists a constant $c>0$ such that $\lambda_{r}(\Sigma) > cp$. \\
(ii) The elements of $B$ are uniformly bounded by a constant $C$.
\end{assmp}

Note first assumption (ii) implies that $\lambda_1(\Sigma) \le \lambda_1(B \Sigma_f B^T) + \|\Sigma_u\| \le \lambda_1(\Sigma_f) \lambda_1(B^T B) + \|\Sigma_u\| = O(p)$. So together with (i), the above assumption requires leading eigenvalues to grow with an order of $p$. This assumption is satisfied by the approximate factor model, since by Weyl's inequality, $\lambda_i(\Sigma)/p = \lambda_i(B\Sigma_f B^T)/p + o(1)$ if the main term is bounded from below. Furthermore, if we assume that each row of $B$ is iid from the same distribution with a finite second moment, it is not hard to see $\lambda_i(B\Sigma_f B^T)/p = \lambda_i(\Sigma_f (B^T B/p)) $ satisfies such a condition. Consequently, it is natural to assume $\lambda_i(\Sigma)/p$ is lower bounded for $i \le r$.

Assumption (ii) is related to the matrix incoherence condition. In fact, when $\lambda_{\text{max}}(\Sigma)$ grows linearly with $p$, the condition that $\| B \|_{\infty}$ is bounded is equivalent to the incoherence of eigenvectors of $\Sigma$ being bounded, which is standard in the matrix completion literature \citep{CanRec09} and the robust PCA literature \citep{Cha11}.

We now consider the moment assumption of random variables in model (\ref{eq1.1}).
\begin{assmp}\label{assmp::moment}
$(f_t, u_t)$ is iid with mean zero and bounded fourth moments. That is, there exists a constant $C>0$ such that $\max_kEf_{kt}^4 < C$ and $\max_iE u_{it}^4 < C$.
\end{assmp}
The independence assumption can be relaxed to mixing conditions, but we do not pursue this direction in the current paper. We are going to establish our results based on the general distribution family with only bounded fourth moment in the above assumption.

\subsection{Robust estimation procedure} \label{sec2.2}
The basic idea we propose is to estimate the covariance matrix of the joint vector $(y_t, f_t)$ instead of just that of $y_t$, although it is our target. The covariance of the concatenated $p+r$ dimensional vector $z_t^T = (y_t^T, f_t^T)$ contains all the information we need to recover the low-ranks and sparse structure. Observe that the covariance matrix $\Sigma_z := \text{Cov}(z_t)$ can be expressed as
\begin{align*}
\Sigma_z &= \left( \begin{array}{cc}
B\Sigma_f B^T + \Sigma_u & B\Sigma_f \\
\Sigma_f B^T & \Sigma_f \end{array} \right)
 =: \left( \begin{array}{cc}
\Sigma_{11} & \Sigma_{12} \\
\Sigma_{21} & \Sigma_{22} \end{array} \right).
\end{align*}
Any method which yields an estimate of $\Sigma_z$ as an initial estimator or estimates of $\widehat{\Sigma}_{11}, \widehat{\Sigma}_{12}, \widehat{\Sigma}_{21}, \widehat{\Sigma}_{22}$ could be used to infer the unknown $B, \Sigma_f$ and $\Sigma_u$. Specifically, using the estimator $\widehat{\Sigma}_z$, we can readily obtain an estimator of $B\Sigma_fB^T$ through the identity
\begin{equation*}
B\Sigma_fB^T = \Sigma_{12}\Sigma_{22}^{-1}\Sigma_{21}.
\end{equation*}
Subsequently, we can subtract the estimator of $B\Sigma_fB^T$ from $\widehat{\Sigma}_{11}$ to obtain $\widehat{\Sigma}_u$. With the sparsity structure of $\Sigma_u$ assumed in Section \ref{sec2.1}, the well-studied thresholding \citep{BicLev08, RotLevZhu09, CaiLiu11} can be employed. Applying thresholding to $\widehat{\Sigma}_u$, we obtain a thresholded matrix $\widehat{\Sigma}_u^{\mathcal{T}}$ with guaranteed error in terms of max norm and operator norm. The final step is to add up $\widehat{\Sigma}_u^{\mathcal{T}}$ with the estimator of  $B\Sigma_fB^T$ from $\widehat{\Sigma}_{z}$ to produce the final $\widehat{\Sigma}^{\mathcal{T}}$ for $\Sigma$.

Due to the fact that we only have bounded fourth moments for factors and errors, a straightforward idea to estimate the covariance matrix $\Sigma_z$ is through robust methodology. For the sake of simplicity, we assume $z_t$ has zero mean, so the covariance matrix of $z_t$ takes the form $Ez_tz_t^T$. We shall use the M-estimator proposed in \cite{Cat12} and \cite{FanLiWan16}, where the authors proved the concentration property in the estimation of population mean of a random variable with only a finite second moment. In essence, minimizing a suitable loss function, say Huber loss, yields an estimator of the population mean with deviation of order $n^{-1/2}$. The Huber loss reads
\begin{equation}
l_{\alpha} (x) = \begin{cases}
2\alpha |x| - \alpha, & |x| > \alpha, \\
x^2, & |x| \le \alpha.
\end{cases}
\end{equation}
Choosing $\alpha = \sqrt{\, (nv^2) /  \log(\epsilon^{-1}) }$, $\epsilon \in (0,1)$ where $v$ is an upper bound of the standard deviation of the random variable $X_i$ of interest, \cite{FanLiWan16} showed that the minimizer $\widehat{\mu} = \argmin_{\mu} \sum_{i=1}^n l_{\alpha}(X_i - \mu)$ satisfies
\begin{equation} \label{eqn::concentration}
P \bigg( |\widehat{\mu} - \mu| \le 4v\sqrt{\frac{\log(\epsilon^{-1})}{n}} \bigg) \ge 1 - 2\epsilon,
\end{equation}
when $n \ge 8 \log(\epsilon^{-1})$ where $\mu = E x_i$. This finite sample result holds for any distributions with bounded second moments, including asymmetric distributions generated by $X = Z^2$.  The diverging parameter $\alpha$ is chosen to reduce the biases of the $M$-estimator for asymmetric distributions and hence we require a finite second moment.  In our covariance matrix estimation, we will take $X_i$ to be the square of a random variable or products of two random variables.  When applying this method to estimate $\Sigma_z$ element-wise, we expect $\widehat\Sigma_{11}, \widehat\Sigma_{12}, \widehat\Sigma_{21},\widehat\Sigma_{22}$ to achieve element-wise errors of $O_P(\sqrt{\log p / n})$, where the logarithmic term is incurred when we bound the errors uniformly. The formal result will be given in Section \ref{sec::theory}.

In an earlier work, \cite{Cat12} proposed solving the equation
$
\sum_{i=1}^n h[ \alpha^{-1}(\mu - \widehat{\mu})] = 0,
$
where the strictly increasing $h(x)$ satisfies $-\log (1-x+x^2/2) \le h(x) \le \log(1+x+x^2/2) $. For $\epsilon \in (0,1)$ and $n > 2\log(\epsilon^{-1})$, \cite{Cat12} proved that
\begin{equation*}
P\bigg( | \widehat{\mu} - \mu | \le v \sqrt{\frac{2 \log(\epsilon^{-1})}{n - 2\log(\epsilon^{-1})}} \bigg) \ge 1 -2 \epsilon,
\end{equation*}
when $n \ge 4\log(\epsilon^{-1})$ and
$
\alpha = \sqrt{{nv^2(1+ \frac{2\log(\epsilon^{-1})}{n-2\log(\epsilon^{-1})})}/\{2\log(\epsilon^{-1})}\},
$
where $v$ is an upper bound of the standard deviation. This $M$-estimator can also be used for covariance estimation, though it usually has a larger bias as shown in \cite{FanLiWan16}.

The whole procedure can be presented in the following steps:
\begin{itemize}[leftmargin= .6in]
\item[\textbf{Step} 1]  For each entry of the covariance matrix $\Sigma_z$, obtain a robust estimator by solving a convex minimization problem (through, for example, Newton-Rapson method):
\begin{equation*}
(\widehat{\Sigma}_z^R)_{ij} = \argmin_x \sum_{t=1}^n l_{\alpha} (z_{it}z_{jt}- x),
\end{equation*}
where $\alpha$ is chosen as discussed above and
$
\widehat\Sigma_z = \widehat\Sigma_z^R =
\left( \begin{array}{cc}
\widehat\Sigma_{11} & \widehat\Sigma_{12} \\
\widehat\Sigma_{21} & \widehat\Sigma_{22} \end{array} \right).
$
\item[\textbf{Step} 2] Derive an estimator of $\Sigma_u$ through the algebraic manipulation
\begin{equation*}
\widehat{\Sigma}_u = \widehat{\Sigma}_{11} - \widehat{\Sigma}_{12} \widehat{\Sigma}_{22}^{-1} \widehat{\Sigma}_{21},
\end{equation*}
and then apply adaptive thresholding of \cite{CaiLiu11}. That is,
\begin{equation*}
(\widehat{\Sigma}_u^{\mathcal{T}})_{ij} = \begin{cases}
(\widehat{\Sigma}_u)_{ij}, & i = j \\
s_{ij}((\widehat{\Sigma}_u)_{ij}) \mathbbm{1}(| (\widehat{\Sigma}_u)_{ij} | \ge \tau_{ij}), & i \neq j
\end{cases}
\end{equation*}
where $s_{ij}(\cdot)$ is a the generalized shrinkage function \citep{AntFan01, RotLevZhu09} and $\tau_{ij} = \tau ((\widehat{\Sigma}_u)_{ii} (\widehat{\Sigma}_u)_{jj})^{1/2}$ is an entry-dependent threshold.
\item[\textbf{Step} 3] Produce the final estimator for $\Sigma$:
\begin{equation*}
\widehat{\Sigma}^{\mathcal{T}} =  \widehat{\Sigma}_{12} \widehat{\Sigma}_{22}^{-1} \widehat{\Sigma}_{21} + \widehat{\Sigma}^{\mathcal{T}}_u.
\end{equation*}
\end{itemize}
Note in the above steps, the choice of the parameters $v$ (in the definition of $\alpha$) and $\tau_{ij}$ are not yet specified and will be discussed in Section \ref{sec::theory}.

Before delving into the analysis of the procedure, we first deviate to look at a technical issue.
Recall that $\widehat{\Sigma}_{22}$ is an estimator of $\Sigma_f$, by Weyl's inequality,
\begin{equation*}
|\lambda_i(\widehat{\Sigma}_{22}) - \lambda_i(\Sigma_f)| \le \| \widehat{\Sigma}_{22} - \Sigma_f \|,
\end{equation*}
Since both matrices are of low dimensionality, as long as we are able to estimate every entry of $\Sigma_f$ accurate enough (see Lemma \ref{bound::robust} below), $\| \widehat{\Sigma}_{22} - \Sigma_f \|$ vanishes with high probability as $n$ diverges. Since $\widehat{\Sigma}_{22}$ is invertible with high probability, there is no major issue implementing the procedure. In cases where positive semidefinite (psd) matrix is expected, we replace the matrix with its nearest positive semidefinite version. We can do this projection for either $\widehat{\Sigma}_u$ or $\widehat{\Sigma}_z$. For example, for $\widehat{\Sigma}_u$, we solve the following optimization problem:
\begin{equation}\label{opt::nearPD}
\widetilde{\Sigma}_u = \argmin_{\Sigma_u \succeq 0} \|\widehat{\Sigma}_u - \Sigma_u \|_{\infty}\,,
\end{equation}
and simply employ $\widetilde{\Sigma}_u$ as a surrogate of $\widehat{\Sigma}_u$.
Observe that
\begin{equation*}
\| \widetilde{\Sigma}_u - \Sigma_u \|_{\infty} \le \| \widetilde{\Sigma}_u - \widehat{\Sigma}_u \|_{\infty}  + \| \widehat{\Sigma}_u - {\Sigma}_u \|_{\infty} \le 2\| \widehat{\Sigma}_u - {\Sigma}_u \|_{\infty}.
\end{equation*}
Thus, apart from a different constant, $\widetilde{\Sigma}_u$ inherits all the desired properties of $\widehat{\Sigma}_u$, and we are able to safely replace $\widehat{\Sigma}_u$ with $\widetilde{\Sigma}_u$ without modifying our estimation procedure. Moreover, (\ref{opt::nearPD}) can be cast into the semidefinite programming problem below,
\begin{equation}\label{opt::SDP}
\min_{t, \Sigma_u \succeq 0} t \;\; \text{s.t.} \;\; |\widehat\Sigma_u - \Sigma_u|_{ij} \le t\,,
\end{equation}
which can be easily solved by a semidefinite programming solver, e.g. \cite{GraBoyYe08}.

\section{Theoretical analysis}\label{sec::theory}

In this section, we will show the theoretical properties of our robust estimator under bounded fourth moments. We will also show that when the data are known to be generated from more restricted families (e.g. sub-Gaussian), commonly used estimators such as sample covariance estimator suffices as an initial estimator in Step $1$.

\subsection{General theoretical properties}

From the above discussion on M-estimators and their concentration results, it is immediate to have the following lemma.

\begin{lem}\label{bound::robust}
Suppose that a $d$-dimensional random vector $X$ is centered and has finite fourth moments, i.e.\ $EX = 0$, $\max_iEX_i^4 < + \infty $ for $i = 1,2,\ldots,p$. Letting $\sigma_{ij} = E(X_iX_j)$ and $\widehat{\sigma}_{ij}$ be Huber's estimator with parameter $\alpha = \sqrt{\, nv^2 / \log(p^2/\delta)}$, then there exists a universal constant $C$ such that for any $\delta \in (0,1)$ and $ n \ge C \log(p/\delta)$, with probability $ 1- \delta$,
\begin{equation}
\max_{ij} \left| \widehat{\sigma}_{ij} - \sigma_{ij} \right| \le C v \sqrt{ \frac{ \log p + \log (1/\delta) }{n}},
\end{equation}
where $v$ is a pre-determined parameter satisfying $v^2 \ge \max_{i,j \le p} \Var (X_iX_j)$.
\end{lem}
In practice, we do not know any of the fourth moments in advance. To pick up a good $v$, one possibility is to try a sequence of geometrically increasing $v$, as studied in \cite{Cat12}. Similar to \cite{FanLiuWan15}, we may also use empirical variance to give a rough bound of $v$.

Recall that $z_t$ is a $p+r$ dimensional vector concatenating $y_t$ and $f_t$. From Assumption \ref{assmp::moment}, there is a constant $C_0$ as a uniform bound for $Ez_{it}^4$. This leads to the following result.
\begin{cor}\label{cor::Sigmaz}
Suppose that $\widehat{\Sigma}_z$ is an estimator of covariance matrix $\Sigma_z$, whose entries are Huber's estimators with parameter $\alpha = \sqrt{\, nv^2 / \log((p+r)^2/\delta)}$. Then there exists a universal constant $C$ such that for any $\delta \in (0,1)$ and $ n \ge C \log(p/\delta)$, with probability $1-\delta$,
\begin{equation}\label{ineqn::concentrate}
\| \widehat{\Sigma}_z - \Sigma_z \|_{\infty} \le Cv\sqrt{\frac{\log p + \log(1/\delta)}{n}},
\end{equation}
where $v$ is a pre-determined parameter satisfying $v^2 \ge C_0$.
\end{cor}

So after Step 1 of the proposed procedure, we obtain an estimator $\widehat\Sigma_z$ that achieves optimal rate of element-wise convergence. With $\widehat\Sigma_z$, we proceed to establish results of estimation errors of our concern.
We will establish convergence rates for both $\widehat{\Sigma}_u^{\mathcal{T}}$ and $\widehat{\Sigma}^{\mathcal{T}}$. The key theorem that links the estimation error under element-wise max norm with that under other norms is stated as follows.

\begin{thm}  \label{thm3.1}
Under Assumptions \ref{assmp::Sigu} - \ref{assmp::moment}, if we have estimator $\widehat{\Sigma}_z$ satisfying
\begin{align}
\| \widehat{\Sigma}_z - \Sigma_z \|_{\infty} = O_P(\sqrt{\log p / n}),
\end{align}
then the three-step procedure in Section \ref{sec2.2} with $\tau \asymp \sqrt{\log p/n}$ generates $\widehat\Sigma_u^{\mathcal{T}}$ and $\widehat\Sigma^{\mathcal{T}}$ satisfying
\begin{equation} \label{eq3.4}
\|\widehat{\Sigma}_u^{\mathcal{T}} - \Sigma_u \|_2 = \|(\widehat{\Sigma}_u^{\mathcal{T}})^{-1} - \Sigma_u^{-1} \|_2 = O_P\Big(m_p \Big(\frac{\log p}{n}\Big)^{(1-q)/2}\Big),
\end{equation}
and furthermore
\begin{align}
&\| \widehat{\Sigma}^{\mathcal{T}} - \Sigma \|_{\infty} = O_P\Big(\sqrt{\frac{\log p}{n}}\Big),\label{eq3.5}\\
&\| \widehat{\Sigma}^{\mathcal{T}} - \Sigma \|_{\Sigma} = O_P\Big(\frac{\sqrt{p}\log p}{n} + m_p\Big(\frac{\log p}{n}\Big)^{(1-q)/2} \Big), \label{eq3.6}\\
&\| (\widehat{\Sigma}^{\mathcal{T}})^{-1} - \Sigma^{-1} \| = O_P\Big(m_p \Big(\frac{\log p}{n}\Big)^{(1-q)/2}\Big), \label{eq3.7}
\end{align}
where $\|A\|_{\Sigma} = p^{-1/2} \|\Sigma^{-1/2} A \Sigma^{-1/2}\|_F$ is the relative Frobenius norm defined in \cite{FanFanLv08}, if $n$ is large enough so that $m_p (\log p/n)^{(1-q)/2}$ is bounded.
\end{thm}


Note that this theorem provides a nice interface that connects max-norm guarantee with the desired convergence rate . Therefore, any robust method that attains the element-wise optimal convergence rate as in Corollary \ref{cor::Sigmaz} can be used in Step 1 instead of the current M-estimator approach.

\subsection{Estimators under more restricted distributional assumptions}

We analyzed theoretical properties of the robust procedure in the previous subsection under the assumption of only bounded fourth moments. Theorem \ref{thm3.1} shows that any estimator that achieves the optimal max norm convergence rate could serve as an initial pilot estimator for $\Sigma_z$ to be used in Step 2 and Step 3 of our procedure. Thus the procedure depends on the distribution assumption \ref{assmp::moment} only through Step 1 where a proper estimator $\widehat\Sigma_z$ is proposed. Sometimes, we do have more information on the shape of the distributions of factors and noises. For example, if the distribution of $z_t = (f_t^T, u_t^T)^T$ has a sub-Gaussian tail, the sample covariance matrix $\widehat\Sigma_z^S = n^{-1} \sum_{t = 1}^n z_t z_t^T$ attains the optimal element-wise maximal rate for estimating $\Sigma_z$.

In an earlier work, \cite{FanLiaMin11} proposed to simply regress observations $y_t$ on $f_t$ in order to obtain
\begin{equation}
\widehat{B} = Y^TF(F^TF)^{-1},
\end{equation}
where $Y = (y_1, \dots, y_n)^T$ and $F = (f_1, \dots, f_n)^T$. Then they threshold the matrix $\widehat\Sigma_u = \widehat\Sigma - \widehat B \widehat \Sigma_f \widehat B^T$ where $\widehat\Sigma = n^{-1} Y Y^T$ and $\widehat \Sigma_f = n^{-1} F^T F$. This regression procedure is equivalent to applying $\widehat\Sigma_z^S$ directly in Step 1 and also equivalent to solving a least-square minimization problem, and thus suffers from robustness issue when the data come from heavy-tailed distributions. All the convergence rates achieved in Theorem \ref{thm3.1} are identical with \cite{FanLiaMin11} where sub-Gaussian tails are assumed.

As we explained, if $z_t$ is sub-Gaussian distributed, $\widehat\Sigma_z^S$ instead of $\widehat\Sigma_z^R$ can be used. If $f_t$ and $u_t$ exhibit heavy tails, another widely used assumption is t-distribution, which is included in the elliptical distribution family. The elliptical distribution is defined as follows. Let $\mu \in \mathbb R^p$ and $\Sigma \in \mathbb R^{p \times p}$ with $\rank(\Sigma) = q \le p$. A $p$-dimensional random vector $y$ has an elliptical distribution, denoted by $y \sim ED_p(\mu, \Sigma, \zeta)$, if it has a representation \citep{FanKotNg90}
\beq \label{Eq:ellip}
y \overset{d} = \mu + \zeta A U\,,
\eeq
where $U$ is a uniform random vector on the unit sphere in $\mathbb R^q$, $\zeta \ge 0$ is a scalar random variable independent of $U$, $A \in \mathbb R^{p \times q}$ is a deterministic matrix satisfying $A A' = \Sigma$. To make the representation \eqref{Eq:ellip} identifiable, we require $\mathbb E\zeta^2 = q$ so that $\Cov(y) = \Sigma$. Here we also assume continuous elliptical distributions with $\mathbb P(\zeta = 0) = 0$.

If $f_t$ and $u_t$ are uncorrelated and jointly elliptical, i.e., $z_t = (f_t^T, u_t^T)^T \sim ED_p(0, \diag(\Sigma_f, \Sigma_u), \zeta)$, then a well known good estimator for the correlation matrix $R$ of $z_t$ is marginal Kendall's tau. Kendall's tau correlation coefficient is defined as
\beq
\hat\tau_{jk} := \frac{2}{n(n-1)} \sum_{i < i'} \sgn((z_{ij} - z_{i'j})(z_{ik} - z_{i'k})) \,,
\eeq
whose population counterpart is
\beq
\tau_{jk} := \mathbb P((z_{1j} - z_{2j})(z_{1k} - z_{2k}) > 0) -  \mathbb P((z_{1j} - z_{2j})(Y_{1k} - Y_{2k}) < 0)\,.
\eeq
For the elliptical family, the key identity $r_{jk} = \sin(\pi \tau_{jk}/2)$ relates Pearson correlation with Kendall's correlation \citep{FanKotNg90}. Using $\hat r_{jk} = \sin (\pi \hat\tau_{jk} /2)\,$, \cite{HanLiu14} showed that $\widehat{R}$ is an accurate estimate of $R$, achieving $\|\widehat R - R\|_{\infty} = O_P(\sqrt{\log p/n})$.
Let $\Sigma_z = D R D$ where $R$ is the correlation matrix and $D = \diag(\sigma_1,\dots, \sigma_p)$ is a diagonal matrix consisting of standard deviations for each dimension. We construct $\widehat\Sigma_z^K$ by separately estimating  $D$ and $R$. As before, if fourth moment exists, we estimate $D$ by only considering $i = j$ in Step 1.

Therefore, if $z_t$ is elliptically distributed, $\widehat\Sigma_z^K$ can be used as the initial pilot estimator for $\Sigma_z$ in Step 1. Note that $\widehat\Sigma_z^K$ is much more computationally efficient than $\widehat\Sigma_z^R$. However, for general heavy-tailed distributions, there is no simple way to connect the usual correlation with Kendall's correlation. Thus we should favor $\widehat\Sigma_z^R$ instead. We will compare the three estimators $\widehat\Sigma_z^S$, $\widehat\Sigma_z^K$ and $\widehat\Sigma_z^R$ throughly through simulations in Section \ref{sec::sim}.

\subsection{Asymptotics of robust mean estimators}\label{sec::asmp}
In this section we look further into robust mean estimators. Though the result we shall present is asymptotic and not essential for our main theorem \ref{thm3.1}, it is interesting in its own right and deserves some treatment.

Perhaps the best known result of Huber's mean estimator is the asymptotic minimax theory. In \cite{Hub64}, Huber considered the so-called $\epsilon$-contamination model:
\begin{equation*}
\mathcal{P}_{\epsilon} = \{ F \,|\, F(x) = (1-\epsilon)G\left(x - \theta \right) + \epsilon H(x), \, H \in \mathcal{F}, \theta \in \mathbf{R}  \},
\end{equation*}
where $G$ is a known distribution, $\epsilon$ is fixed and $\mathcal{F}$ is the family of symmetric distributions. Let $T_n$ be the minimizer of $\sum_{i=1}^n \rho_H(x_i - \mu)$, where $\rho_H(x) = x^2/2$ for $|x| < \alpha$, and $\rho_H(x) = \alpha|x| - \alpha^2/2$ for $|x| \ge \alpha$, where $\alpha$ is fixed.  In the special case where $G$ is Gaussian, Huber's result shows that with appropriate choice of $\alpha$, Huber's estimator minimizes the maximal asymptotic variance among all translation invariant estimators, the maximum being taken over $\mathcal{P}_{\epsilon}$.

One problem with $\epsilon$-contamination model is that it makes sense only when we assume symmetry of $H$, if $\theta$ is the quantity we are interested in. In contrast, \cite{Cat12} and \cite{FanLiWan16} studied a different family, in which distributions have finite second moments. \cite{Bic76} called them `local' and `global' models respectively, and offered a detailed discussion.

This paper, along with the preceding two papers \citep{Cat12,FanLiWan16}, studies robustness in the sense of the second model. The technical novelty primarily lies in the nice concentration property, which is a powerful tool in high dimensional statistics. This requires the parameter $\alpha$ of $\rho_H$ to grow with $n$, versus being kept fixed, such that the condition in Corollary $\ref{cor::Sigmaz}$ is satisfied. It turns out that, in addition to the concentration property, we can establish results regarding its asymptotic behaviors in an exact manner.

Let $\rho_n(x) = x^2/2$ for $|x| < \alpha_n$ and $\rho_n(x) = \alpha_n|x| - \alpha_n^2/2$ for $|x| \ge \alpha_n$; its derivative $\psi_n = \rho_n'$. Let us write $\lambda_n(t) = E\psi_n(X - t)$. Denote $t_n$ as a solution of $\lambda_n(t) = 0$, which is unique when $n$ is sufficiently large, and $T_n$ a solution of $\sum_{i=1}^n \psi_n(x_i - t) = 0$. We have the following theorem. 

\begin{thm}\label{thm:asmp}
Suppose that $x_1,\ldots, x_n$ is drawn from some distribution $F$ with mean $\mu$ and finite variance $\sigma^2$. Suppose $\{ \alpha_n \}$ is any sequence with $\lim_{n \to \infty} \alpha_n = \infty$. Then, as $n \to \infty$,
\begin{equation*}
\sqrt{n}\, (T_n - t_n) \xrightarrow{d} N(0, \sigma^2),
\end{equation*}
and moreover
\begin{equation*}
\frac{ t_n - \mu}{E \psi_n(X - \mu)} \to 1.
\end{equation*}
\end{thm}

This theorem gives a decomposition of error $T_n - \mu$ into two components: variance and bias. The rate of bias
$E \psi_n(X - \mu)$ depends on the distribution $F$ and $\{ \alpha_n \}$. When the distribution is either symmetric or $\liminf_n \alpha_n / \sqrt{n} > 0$, the second component $t_n - \mu$ is $ o(1/\sqrt{n})$, a negligible quantity compared with the asymptotic variance. While Huber's approach needs the symmetric restriction, there is no need for our estimator. This theorem also lends credibility to the bias-variance tradeoff we observed in the simulation (see Section \ref{sec::tradeoff}).

It is worth comparing the above Huber loss minimization with another candidate for robust mean estimation called ``median-of-means'' estimator given by \cite{HsuSab14}. The method, as its name suggests, first divides samples into $k$ subgroups and calculates means for each subgroup, then take the median of those means as the final estimator. The first step basically symmetrizes the distribution by the central limit theorem and the second step is to robustify the procedure.  According to \cite{HsuSab14}, if we choose $k = 4.5 \log(p/\delta)$ and element-wisely estimate $\Sigma_z$, similar to (\ref{eqn::concentration}), with probability $1-\delta$, we have
$$
\| \widehat{\Sigma}_z - \Sigma_z \|_{\infty} \le 3\sqrt{3}v \sqrt{\frac{\log p + \log(1/\delta)}{n}}.
$$
Although ``median-of-means'' has the desired concentration property,
unlike our estimator here, its asymptotic behavior differs from the empirical mean estimator, and as a consequence, it is not asymptotically efficient when the distribution $F$ is Gaussian. Therefore, regarding efficiency, we prefer our proposed procedure in Section \ref{sec2.2}.

\section{Simulations}\label{sec::sim}

We now present simulation results to demonstrate improvement of the proposed robust method over the least-square based method \citep{FanFanLv08, FanLiaMin11} and Kendall's tau based method \citep{HanLiu14, FanLiuWan15} when factors and errors are heavy-tailed and even elliptically distributed.

However, one must be cautious of the choice of the tuning parameter $\alpha$, since it plays an important role in the quality of the robust estimates. Out of this concern, we shall discuss the intricacy of choosing parameter $\alpha$ before presenting the performance of robust estimates of covariance matrices.

\subsection{Robust estimates of variances and covariances}\label{sec::tradeoff}

For random variables $X_1, \ldots, X_p$ with zero mean that may potentially exhibits heavy-tailed behavior, the sample mean of $v_{ij} = E(X_iX_j)$ is not good enough for our estimation purpose. Though being unbiased, in the high dimensional setting, there is no guarantee that multiple sample means stay close to the true values simultaneously.

As shown in theoretical analysis, this problem is alleviated for robust estimators constructed through M-estimators, whose influence functions grow slowly at extreme values. The desired concentration property in (\ref{ineqn::concentrate}) depends on the choice of parameter $\alpha$, which decides the range outside which large values cease to become more influential. However, in practice, we have to make a good guess of $\text{Var}(X_iX_j)$ as the theory suggests; even so, we may be too conservative in the choice of $\alpha$.  

\begin{figure}[h!]
\centering
\includegraphics[scale = 0.5]{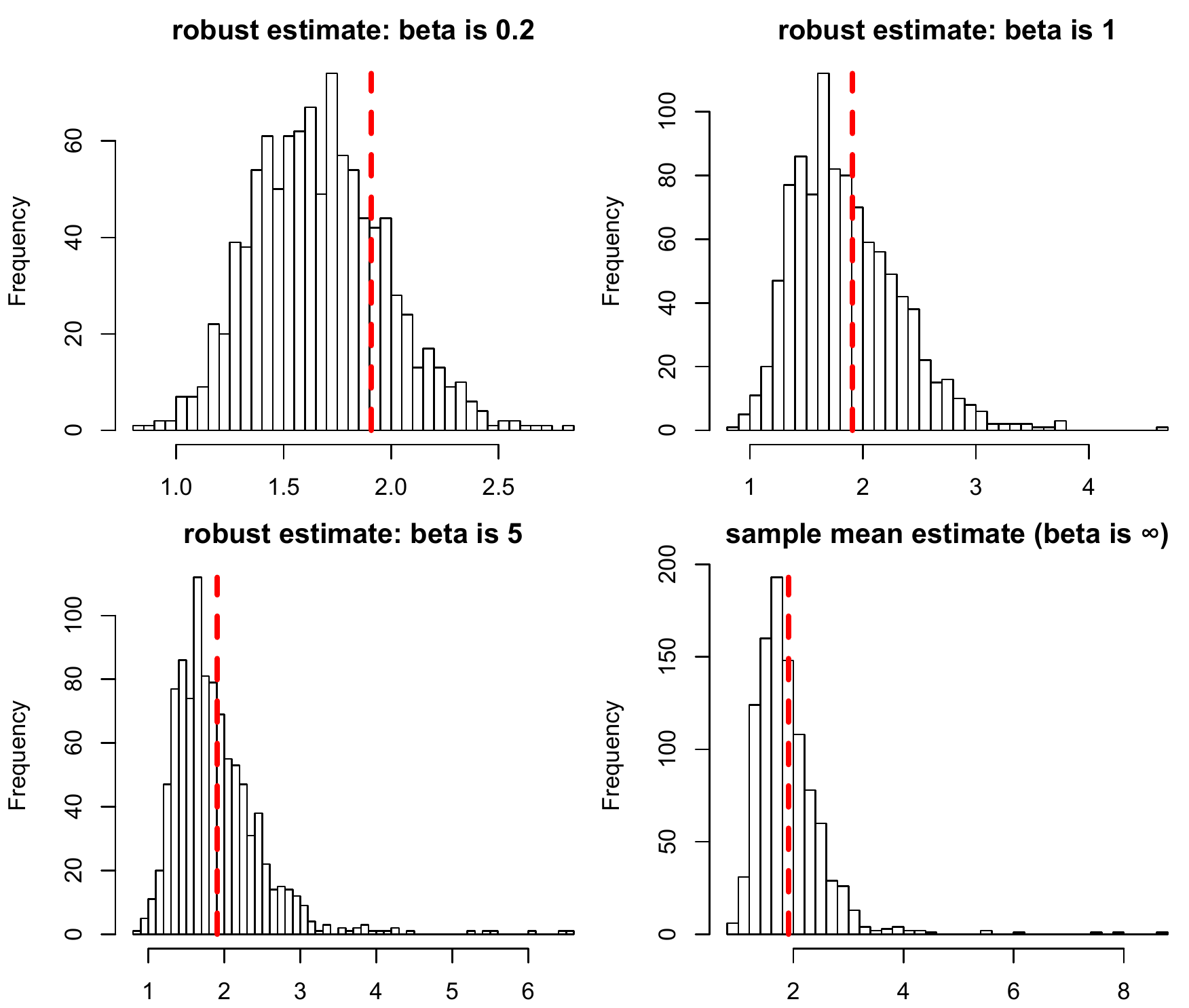}
\caption{\textit{The histograms show the estimates of $\Var(X_i)$ with different paramters $\alpha$ in $1000$ runs. $X_i \sim t_{4}$ so that the true variance $\Var(X_i) = 1.909$. The sample size $n = 100$.}}
 \label{fig::robustVar}
\end{figure}

\begin{figure}[h!]
\centering
\includegraphics[scale = 0.5]{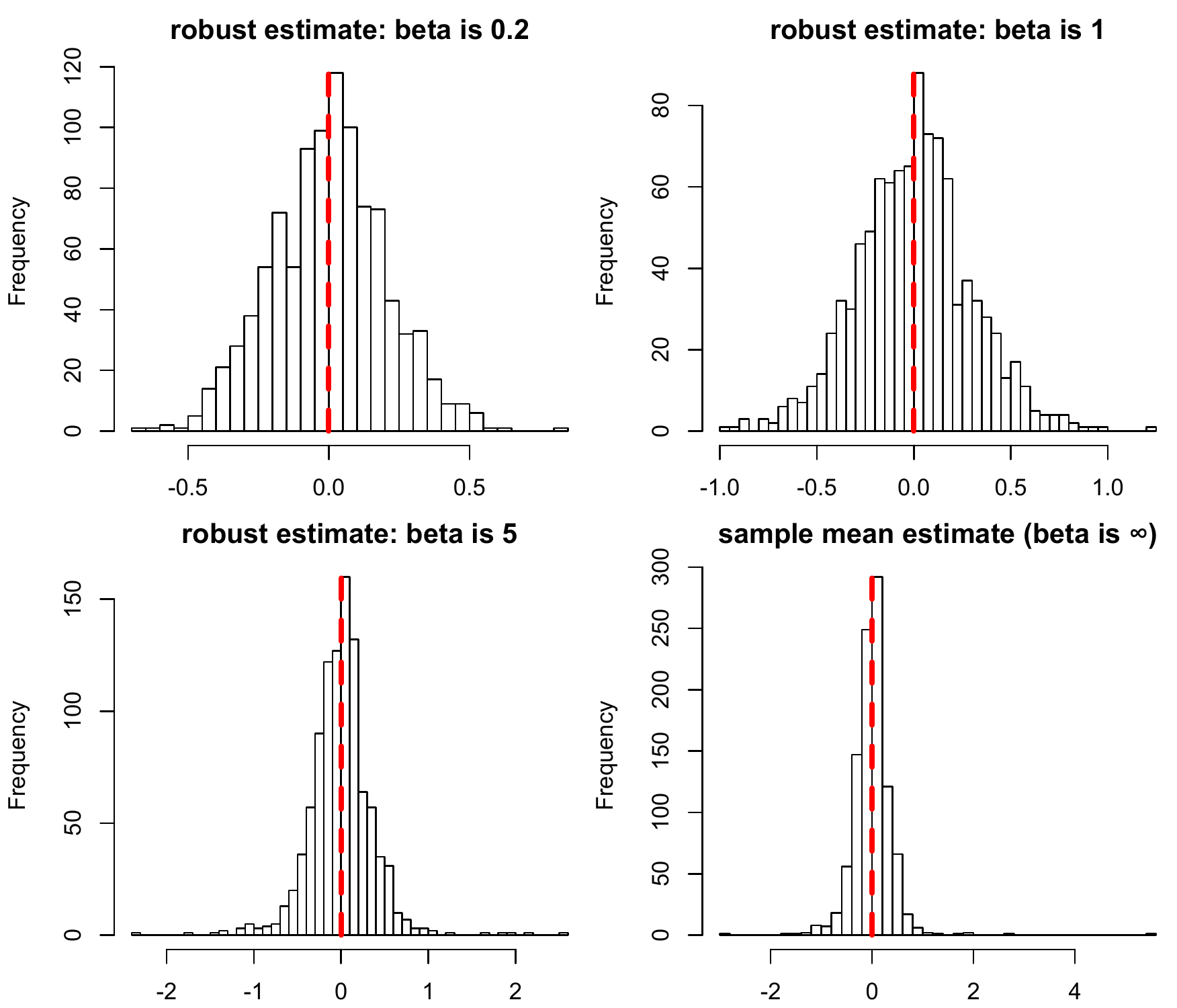}
\caption{\textit{The histograms show the estimates of $ \text{Cov}(X_i,X_j)$ with different paramters $\alpha$ in $1000$ runs. The true covariance $\Cov(X_i,X_j) = 0$. $n = 100$ and the degree of freedom is $4$.}}
 \label{fig::robustCor}
\end{figure}

To show this, we plot in Figure \ref{fig::robustVar} the histograms of our estimates of $v = \text{Var}(X_i)$ in $1000$ runs, where $X_i$ is generated from a t-distribution with degree of freedom $\nu = 4$. The first three histograms show the estimates constructed from Huber's M-estimator, with parameter
\begin{equation}\label{eqn::tuningpar}
\alpha = \beta \sqrt{\frac{n\,\text{Var}(X_i^2)}{2}},
\end{equation}
where $\beta$ is $0.2, 1, 5$ respectively, and the last histogram is the usual sample estimate (or $\beta = \infty$). The quality of estimates ranges from large biases to large variances.  We also plot in Figure \ref{fig::robustCor} the histograms of estimates of $v = \text{Cov}(X_i,X_j)$, where $(X_i,X_j), \; i \neq j$ is generated from a multivariate t-distribution with $\nu = 4$ and an identity scale matrix. The only difference is that in (\ref{eqn::tuningpar}), the variance of $X_i^2$ is replaced by the covariance of $X_iX_j$.

From Figure \ref{fig::robustVar}, we observe a bias-variance tradeoff phenomenon as $\alpha$ varies. This is also consistent with the theory in Section \ref{sec::asmp}. When $\alpha$ is small, the robust method underestimate the variance, yielding a large bias due to the asymmetric of the distribution of $X_i^2$.
As $\alpha$ increases, a larger variance is traded for a smaller bias, until $\alpha = \infty$, in which case the robust estimator simply becomes the sample mean.

For the covariance estimation, Figure \ref{fig::robustCor} exhibits a different phenomenon. Since the distribution of $X_iX_j$ is symmetric for $i \neq j$, there is no bias incurred when $\alpha$ is small. Since the variance is smaller when $\alpha$ is smaller, we have a net gain in terms of the quality of estimates. In the extreme case where $\alpha$ is zero, we are actually estimating the median. Fortunately, under distributional symmetry, the mean and the median are the same.

The simple simulations help us to understand how to choose $\alpha$ in practice: if the distribution is close to a symmetric one, one can choose $\alpha$ aggressively, i.e.\ making $\alpha$ smaller; otherwise, a conservative $\alpha$ is preferred.
 %

\subsection{Covariance matrix estimation}
We implemented the robust estimation procedure with three initial pilot estimators $\widehat\Sigma_z^S$, $\widehat\Sigma_z^K$ and $\widehat\Sigma_z^R$. We simulated $n$ samples of $z_t = (f_t^T, u_t^T)^T$ from a multivariate t-distribution with covariance matrix diag$\{ I_r, 5I_p \}$ and various degrees of freedom. Each row of $B$ is independently sampled from a standard normal distribution. The population covariance matrix of $y_t = B f_t + u_t$ is $ \Sigma = B B^T + 5I_p$. For $p$ running from $200$ to $900$ and $n = p/2$, we calculated errors of the robust procedure in different norms. As suggested by the experiments in the previous section, we chose a larger parameter $\alpha$ to estimate the diagonal elements of $\Sigma_z$, and a smaller one to estimate its off-diagonal elements.
We used the thresholding parameter $\tau = 2 \sqrt{\log p/n}$.

\begin{figure}[h!]
\centering
\includegraphics[scale = 0.55]{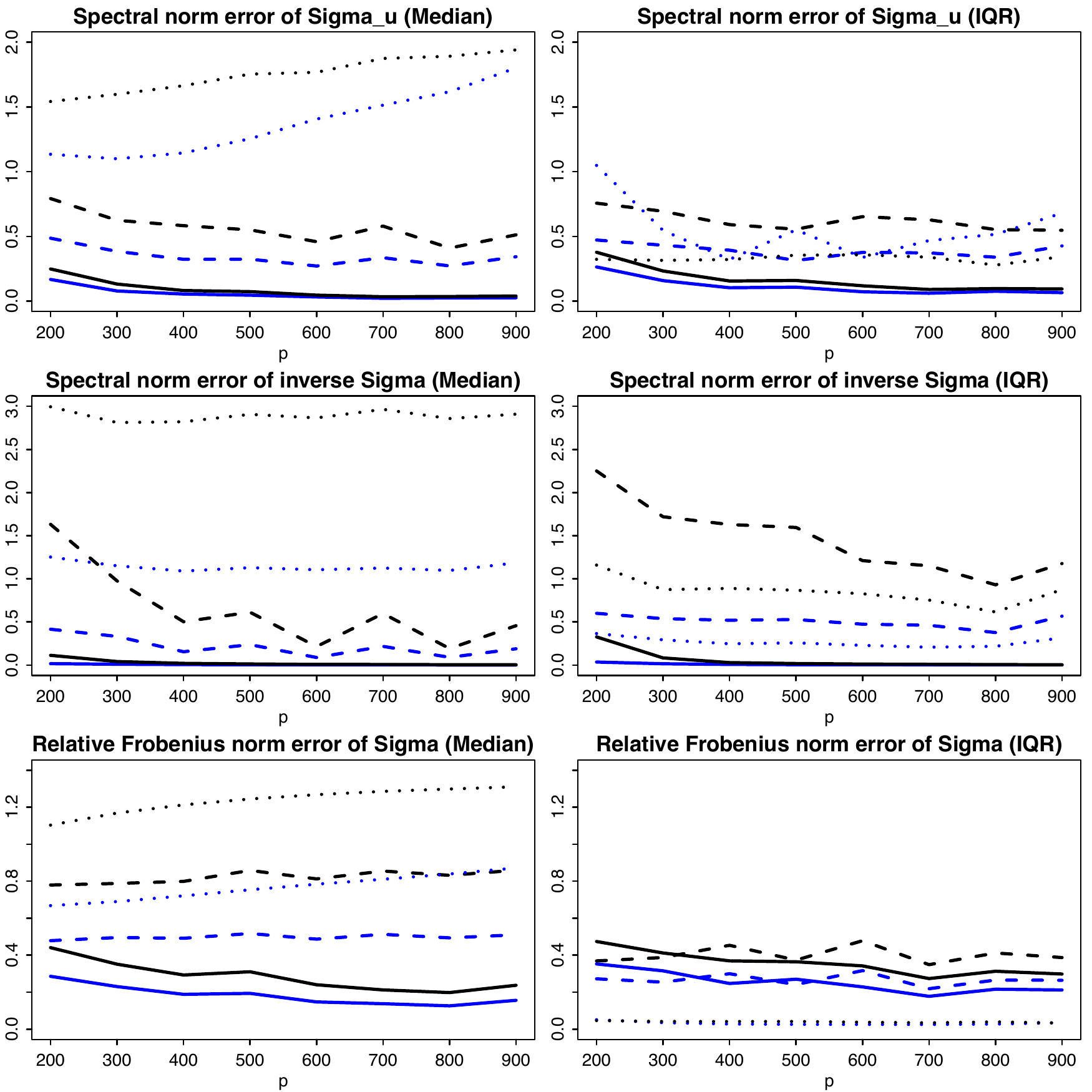}
\caption{\textit{Errors of robust estimates against varying dimensionality $p$. Blue line represents ratio of errors with $\widehat\Sigma_z^R$ over errors with $\widehat\Sigma_z^S$, while black line represents ratio of errors with $\widehat\Sigma_z^K$ over errors with $\widehat\Sigma_z^S$. $z_t$ is generated by multivariate t-distribution with $df = 3$ (solid), $5$ (dashed) and $\infty$ (dotted). The median errors and their IQR over $100$ simulations are reported.}}
\label{figure::exp1}
\end{figure}

\begin{figure}[h!]
\centering
\includegraphics[scale = 0.55]{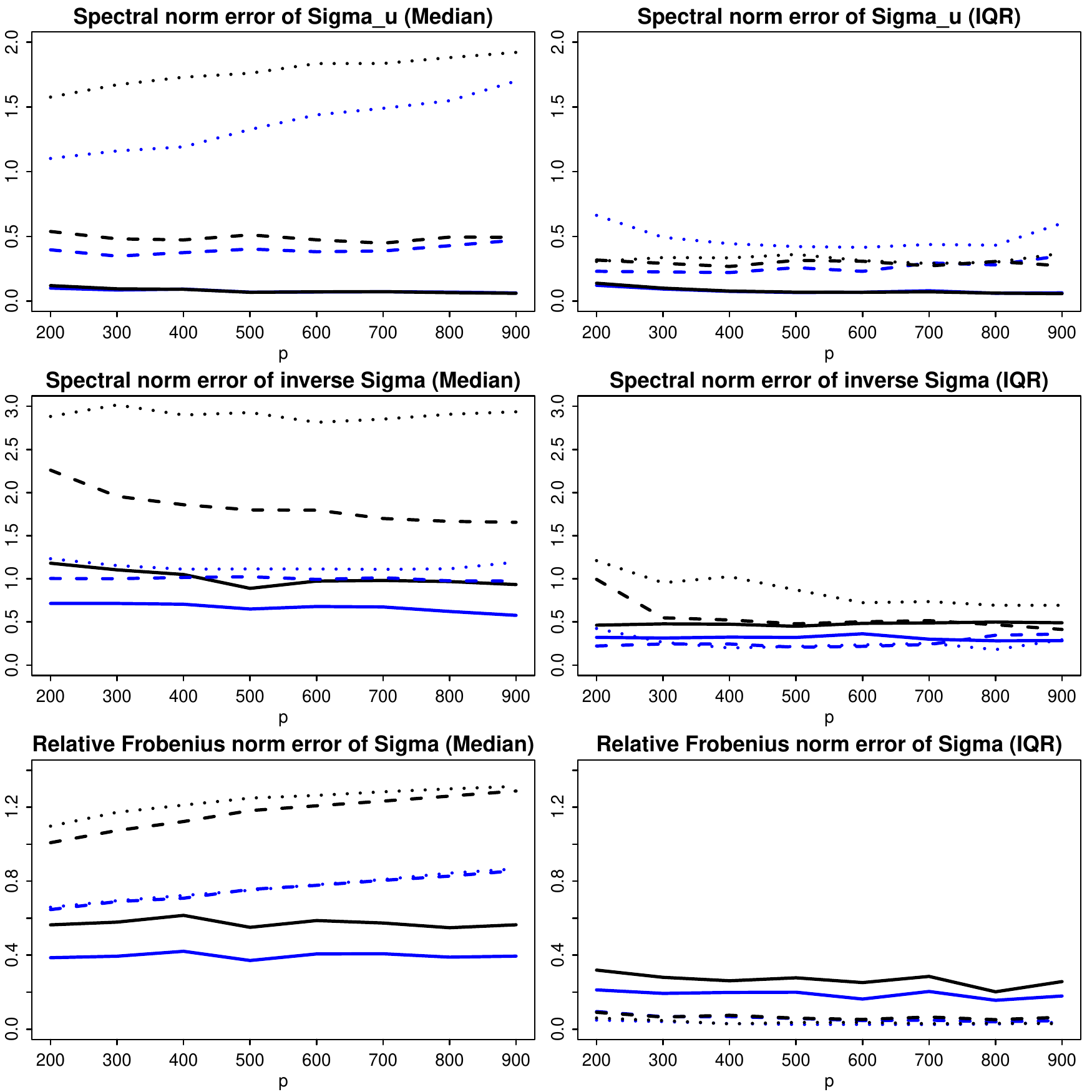}
\caption{\textit{Errors of robust estimates against varying $p$. Blue line represents ratio of errors with $\widehat\Sigma_z^R$ over errors with $\widehat\Sigma_z^S$, while black line represents ratio of errors with $\widehat\Sigma_z^K$ over errors with $\widehat\Sigma_z^S$. $z_t$ is generated by element-wise iid t-distribution with $df = 3$ (solid), $5$ (dashed) and $\infty$ (dotted). The median errors and their IQR over $100$ simulations are reported.}}
\label{figure::exp2}
\end{figure}

The estimation errors are gauged in the following norms: $\| \widehat{\Sigma}_u^{\mathcal{T}} - \Sigma_u \|$, $\| (\widehat{\Sigma}^{\mathcal{T}})^{-1} - \Sigma^{-1} \|$ and $\|\widehat{\Sigma}^{\mathcal{T}} - \Sigma\|_{\Sigma}$ as shown in Theorem \ref{thm3.1}. We considered two different settings: (1) $z_t$ is generated from multivariate t distribution with very heavy ($\nu = 3$), medium heavy ($\nu = 5$), and light ($\nu = \infty$ or Gaussian) tail; (2) $z_t$ is element-wise iid one-dimensional t distribution with degree of freedom $\nu = 3, 5$ and $\infty$. They are separately plotted in Figures \ref{figure::exp1} and \ref{figure::exp2}. The estimation errors of applying sample covariance matrix $\widehat\Sigma_z^S$ are used as the baseline for comparison. For example, if $\|\widehat{\Sigma}^{\mathcal{T}} - \Sigma\|_{\Sigma}$ is used to measure performance, the blue curve represents ratio $\|(\widehat{\Sigma}^{\mathcal{T}})^R - \Sigma\|_{\Sigma}/\|(\widehat{\Sigma}^{\mathcal{T}})^S - \Sigma\|_{\Sigma}$ while the black curve represents ratio $\|(\widehat{\Sigma}^{\mathcal{T}})^K - \Sigma\|_{\Sigma}/\|(\widehat{\Sigma}^{\mathcal{T}})^S - \Sigma\|_{\Sigma}$ where $(\widehat{\Sigma}^{\mathcal{T}})^R, (\widehat{\Sigma}^{\mathcal{T}})^K, (\widehat{\Sigma}^{\mathcal{T}})^S$ are respectively estimators given by the robust procedure with initial pilot estimators $\widehat\Sigma_z^R, \widehat\Sigma_z^K, \widehat\Sigma_z^S$ for $\Sigma_z$. Therefore if the ratio curve moves below $1$, the method is better than naive sample estimator given in \cite{FanLiaMin11} and vice versa. The more it gets below $1$, the more robust the procedure is against heavy-tailed randomness.

The first setting (Figure \ref{figure::exp1}) represents a heavy-tailed elliptical distribution, where we expect the two robust methods work better than the sample covariance based method, especially in the case of extremely heavy tails (solid lines for $\nu = 3$). As expected, both black curves and blue curves under the three measures behave visibly better (smaller than $1$). On the other hand, if data are indeed Gaussian (dotted line for $\nu = \infty$), the method with sample covariance performs better under most measures (greater than $1$). Nevertheless, our robust method still performs comparably with the sample covariance method, as the median error ratio stays around $1$ whereas Kendall's tau method can be much worse than the sample covariance method. A plausible explanation is that the variance reduced compensates for the bias incurred in our procedure. In addition, the IQR plots tell us the proposed robust method is indeed more stable than Kendall's tau.

The second setting (Figure \ref{figure::exp2}) provides an example of non-elliptical distributed heavy-tailed data. We can see that the performance of the robust method dominates the other two methods, which verifies the approach in this paper especially when data comes from a general heavy-tailed distribution. While our method is able to deal with more general distributions, Kendall's tau method does not apply to distributions outside the elliptical family, which excludes the element-wise iid $t$ distribution in this setting. This explains why under various measures, our robust method is better than Kendall's tau method by a clear margin. Note that even in the first setting where the data are indeed elliptical, with proper tuning, the proposed robust methods can still outperform Kendall's tau.

\section{Real data analysis} \label{sec::data}
In this section, we will look into financial historical data during 2005 - 2013, and assess to what extent our factor model characterizes the data.

The dataset we used in our analysis consists of daily returns of $393$ stocks, all of which are large market capitalization constituents of S\&P $500$ index, collected without missing values from 2005 to 2013. This dataset has also been used in \cite{FanLiaWan14}, where they investigated how covariates (e.g.\ size, volume) could be utilized to help estimate factors and factor loadings, whereas the focus of the current paper is to develop robust methods in the presence of heavy tailed data.

In addition, we collected factors data for the same period, where the factors are calculated according to Fama-French three-factor model \citep{FamFre93}. After centering, the panel matrix we will use for analysis, is a $393$ by $2265$ matrix $Y$, in addition to a factor matrix $F$ of size $2265$ by $3$. Here $2265$ is the number of daily returns and $393$ is the number of stocks.

\subsection{Heavy tailedness}
\begin{figure}[h!]
\centering
\includegraphics[scale = 0.55]{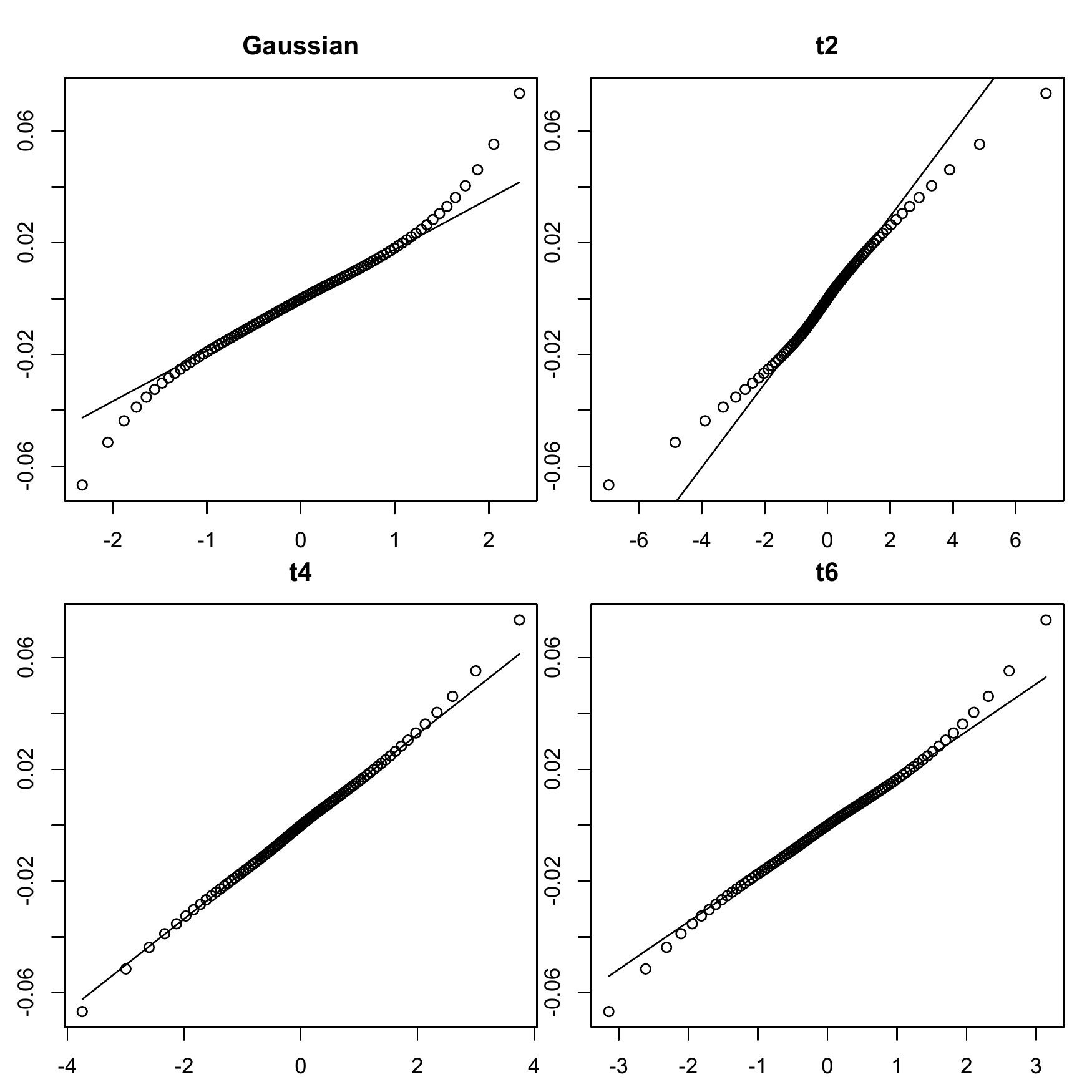}
\caption{\textit{Q-Q plot of excess returns $y_{it}$ for all $i$ and $t$ against Gaussian distribution and t-distribution with degree of freedom $2,4$ and $6$. For each plot, a line is fitted by connecting points at first and third quartile.}}
 \label{fig:return_qq}
\end{figure}

\begin{figure}[h!]
\centering
\includegraphics[scale = 0.55]{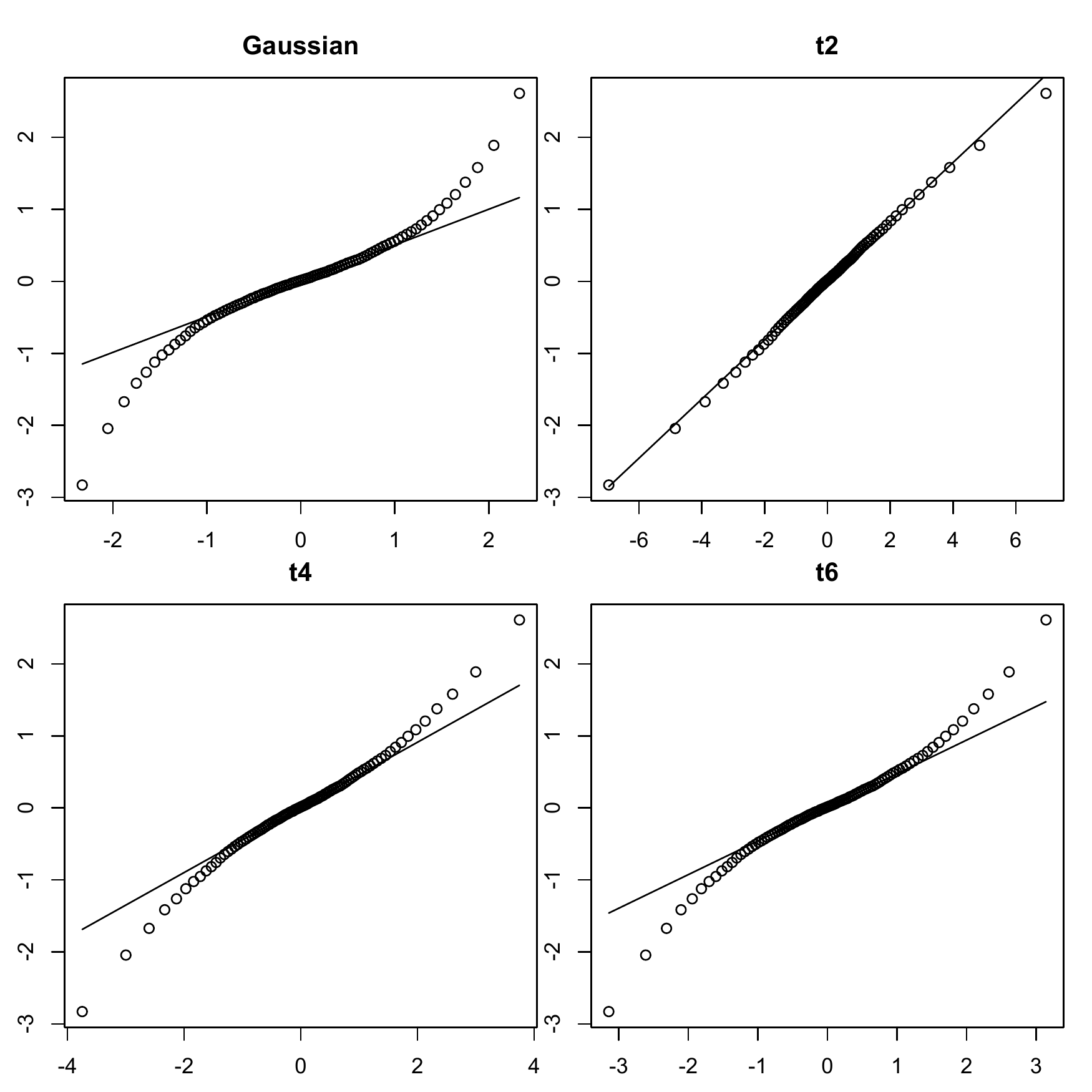}
\caption{\textit{Q-Q plot of factor $f_{it}$ against Gaussian distribution and t-distribution with degree of freedom $2,4$ and $6$. For each plot, a line is fitted by connecting points at first and third quartile.}}
 \label{fig:factor_qq}
\end{figure}

First, we look at how the daily returns are distributed. Especially, we are interested in the behaviors of their tails. In Figure \ref{fig:return_qq}, we made Q-Q plots that compare the distribution of $y_{it}$ with either Gaussian distribution or t-distributions. In the four plots, the base distributions are Gaussian distribution, and t-distribution with varying degree of freedom, ranging from $\text{df}=2$ to $\text{df}=6$. We also fit a line for each plot, showing how much the return data deviate from the base distribution. It is clear that the data has a tail heavier than that of a Gaussian distribution, and that t-distribution with $\text{df} = 4$ is almost in alignment with the return data. Similarly, we made the Q-Q plots for the factors in Figure \ref{fig:factor_qq}. The plots also show that $t$-distribution is better in terms of fitting the data; however, the tails are even heavier, and t-distribution $t_2$ seems to best fit the data.

%

%
%


%
%

\subsection{Spiked covariance structure}

\begin{figure}[h!]
\centering
\includegraphics[scale = 0.37]{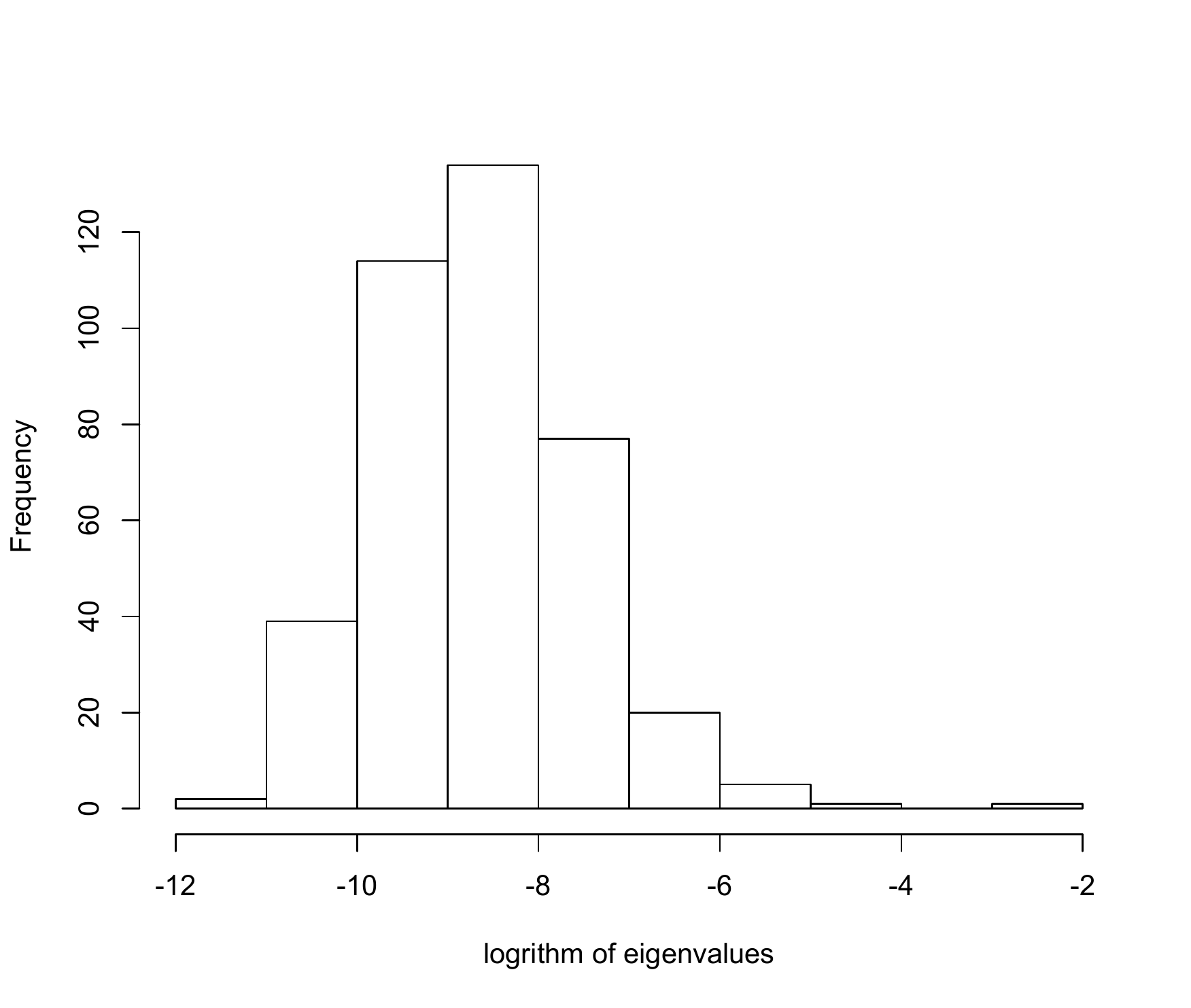} \hspace{0.3 in} \includegraphics[scale = 0.3]{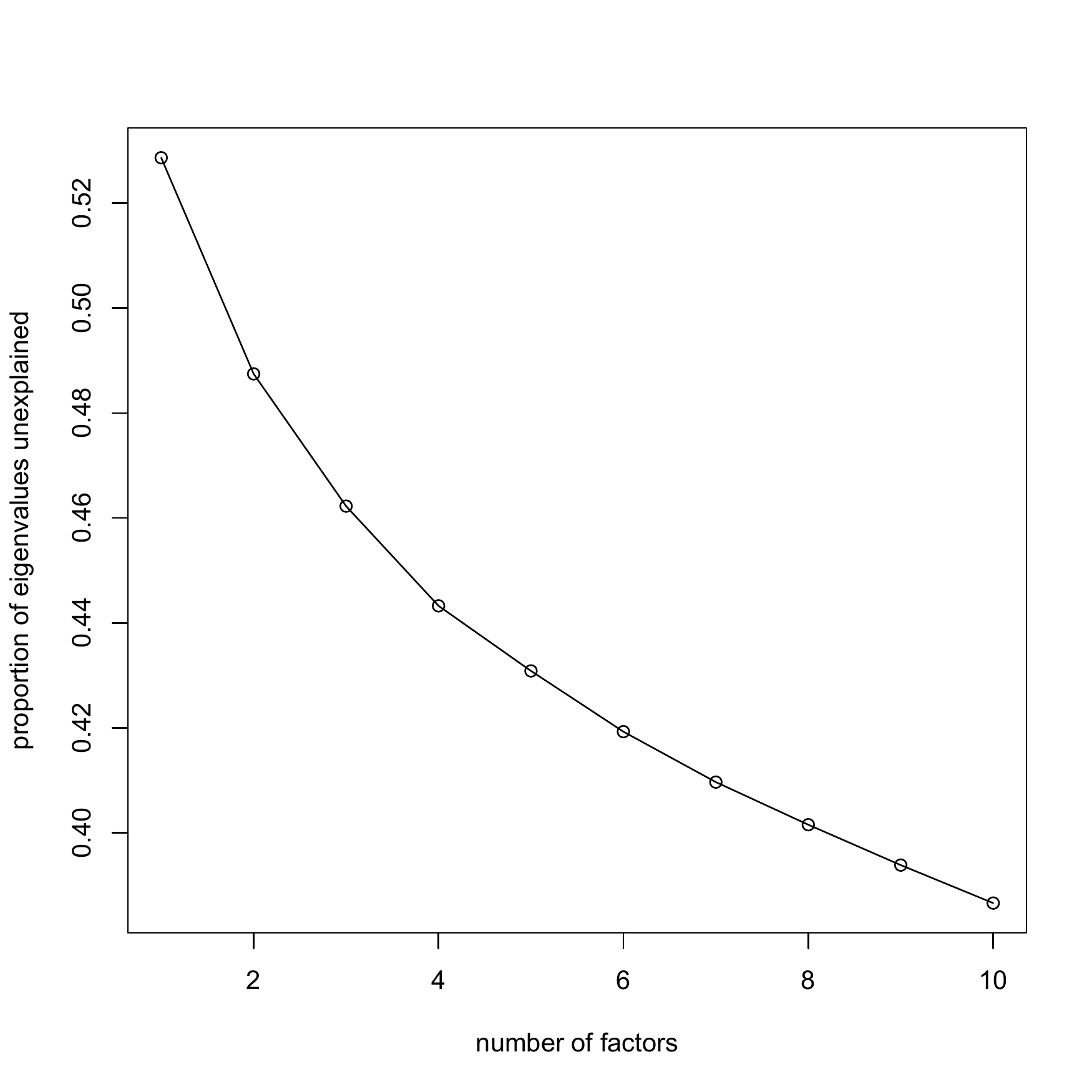}
\caption{\textit{Left panel: Histogram of eigenvalues of sample covariance matrix $Y Y^T/n$. The histogram is plotted on the logarithmic scale, i.e.\ each bin counts the number of $\log \lambda_i$ in a given range. Right panel: Proportion of residue eigenvalues $\sum_{i=K+1}^p \lambda_i / \sum_{i=1}^p \lambda_i$, against varying $K$, where $\lambda_i$ is the $i^{th}$ largest eigenvalue of sample covariance matrix $Y Y^T/n$.  }}
 \label{fig::eigen_hist}
\end{figure}


We now consider how the covariance matrix of returns looks like, since a spiked covariance structure would justify the pervasiveness assumption. To find the spectral structure, we calculated eigenvalues of the sample covariance matrix $ Y Y^T /n$, and made a histogram based on logarithmic scale (see the left panel Figure \ref{fig::eigen_hist}). In the histogram, the counts in the rightmost four bins are $5$, $1$, $0$ and $1$, representing only a few large eigenvalues, which is a strong signal of a spiked structure. We also plotted the proportion of residue eigenvalues $\sum_{i=K+1}^p \lambda_i / \sum_{i=1}^p \lambda_i$, against $K$ in the right panel of Figure \ref{fig::eigen_hist}. The top $3$ eigenvalues account for a major part of the variances, which lends weight to the pervasive assumption.

The spiked covariance structure has been studied in \cite{Pau07}, \cite{JohLu09} and many other papers, but under their regime, the top eigenvalues or ``spiked'' eigenvalues do not grow with the dimension. In this paper, the spiked eigenvalues have stronger signals, and thus are easier to be separated from the rest of eigenvalues. In this respect, the connotation of ``spiked covariance structure'' is closer to that in \cite{FanWan15}. As empirical evidence, this phenomenon also buttresses the motivation of study in \cite{FanWan15}.

\subsection{Portfolio risk estimation}

We consider portfolio risk estimation. To be specific, for a portfolio with weight vector $w \in \mathbf{R}^p$ on all the market assets, its risk is measured by quantity $w^T \Sigma w$ where $\Sigma$ is the true covariance of excess returns of all the assets. Note that $\Sigma$ might be time varying. Here we consider a class of weights with gross exposure $c \ge 1$, that is $\sum_i w_i = 1$ and $\sum_i |w_i| = c$. We consider four scenarios $c = 1, 1.4, 1.8, 2.2$. Note that $c=1$ represents the case of no short selling and the other $c$ values measure different levels of exposure to short selling.

\begin{figure}
\centering
\includegraphics[scale = 0.6]{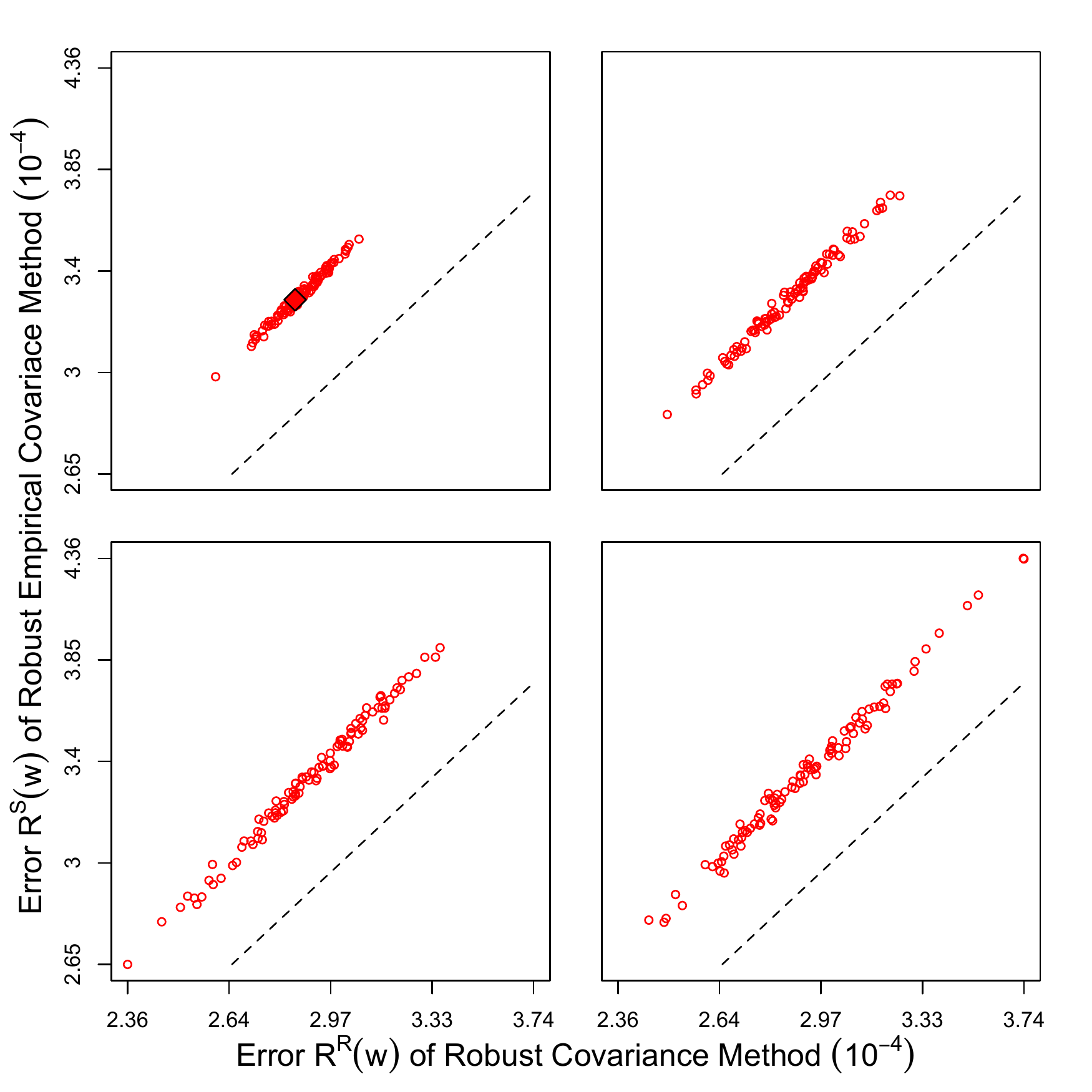}
\caption{\textit{$(R^R(w), R^S(w))$ for multiple randomly generated $w$. The four plots compare the errors of the two methods under different settings (upper left: no short selling; upper right: exposure $c = 1.4$; lower left: exposure $c = 1.8$; lower right: exposure $c = 2.2$). The red diamond in the first plot corresponds to uniform weights. The dashed line is the 45 degree line representing equal performance. Our robust method gives smaller errors.}}
\label{fig:swindow_plot}
\end{figure}

To assess how well our robust estimator performs compared with sample covariance, we calculated the covariance estimators $\widehat{\Sigma}_t^R$ and $\widehat{\Sigma}_t^S$, using the daily data of preceding $12$ months, where $\widehat{\Sigma}_t^R$ is our robust covariance estimator and $\widehat{\Sigma}_t^S$ is the sample covariance, for every trading day from 2006 to 2013. We indexed those dates by $t$ where $t$ runs from $1$ to $2013$. Let $\gamma_t$ be the excess return of the following trading day after $t$. For a weight vector $w$, the error we used to gauge the two approaches is
\begin{equation*}
R^R(w) = \frac{1}{2013}\sum_{t=1}^{2013} \big| w^T \widehat{\Sigma}_t^R w - (w^T \gamma_t)^2 \big|, \quad R^S(w) = \frac{1}{2013}\sum_{t=1}^{2013} \big| w^T \widehat{\Sigma}_t^S w - (w^T \gamma_t)^2 \big|.
\end{equation*}
Note the bias-variance decomposition $E| w^T \widehat{\Sigma}_t w - (w^T \gamma_t)^2 |^2 = E | (w^T \gamma_t)^2 - w^T \Sigma_t w|^2 + E | w^T \widehat{\Sigma}_t w - w^T \Sigma_t w |^2$ where $\Sigma_t = E \gamma_t \gamma_t^T$. The first term measures the systematic risk that cannot be reduced while the second term is the estimation error for the risk of portfolio $w$.  

To generate multiple random weights $w$ with gross exposure $c$, we adopted the strategy used in \cite{EFE15}: (1) for each index $i \le p$ let $\eta_i = 1$ (long) with probability $(c+1)/2c$ and $\eta_i = -1$ (short) with probability $(c-1)/2c$; (2) generate iid $\xi_i$ by exponential distribution; (3) for $\eta_i = 1$, let $w_i = \frac{c+1}{2} \cdot \xi_i / \sum_{\eta_i = 1} \xi_i $ and for $\eta_i = -1$, let $w_i = -\frac{c-1}{2} \cdot \xi_i / \sum_{\eta_i = -1} \xi_i$.
We made a set of scatter plots in Figure \ref{fig:swindow_plot}, in which the x-axis represents $R^R(w)$ and the y-axis $R^S(w)$. In addition, we highlighted in the first plot the point with uniform weights (i.e.\ $w_i = 1/p$), which serves as a benchmark for comparison. The dashed line shows where the two approaches have the same performance. Clearly, for all $w$ the robust approach has smaller risk errors, and therefore has better empirical performance in estimating portfolio risks.

\appendix

\section{Appendix}  \label{sec::appA}

\begin{proof}[\bf Proof of Theorem \ref{thm3.1}]
Since we have robust estimator $\widehat\Sigma_z$ such that $\|\widehat\Sigma_z - \Sigma_z\|_{\infty} = O_P(\sqrt{\log p/n})$, we clearly know $\widehat\Sigma_{11}, \widehat\Sigma_{12}, \widehat\Sigma_{21}, \widehat\Sigma_{22}$ achieve the same rate. Using this, let us first prove $\|\widehat\Sigma_u - \Sigma_u\|_{\infty} = O_P(\sqrt{\log p/n})$. Obviously,
\beq
\|\widehat\Sigma_{12}\widehat\Sigma_{22}^{-1}\widehat\Sigma_{21}^T - B\Sigma_f B^T\|_{\infty} = \|\widehat\Sigma_{12}\widehat\Sigma_{22}^{-1}\widehat\Sigma_{21}^T - \Sigma_{12} \Sigma_{22}^{-1} \Sigma_{21}^T\|_{\infty} = O_P(\sqrt{\log p/n})\,,
\eeq
because the multiplication is along the fixed dimension $r$ and each element is estimated with the rate of convergence $O_P(\sqrt{\log p/n})$. Also $\|\widehat\Sigma_{11} - \Sigma\|_{\infty} = O_P(\sqrt{\log p/n})$, therefore $\widehat\Sigma_u = \widehat\Sigma_{11} - \widehat\Sigma_{12}\widehat\Sigma_{22}^{-1}\widehat\Sigma_{21}^T$ is good enough to estimate $\Sigma_u = \Sigma - B\Sigma_f B^T$ with error $O_P(\sqrt{\log p/n})$ in max norm.

Once the max error of sparse matrix $\Sigma_u$ is controlled, it is not hard to show the adaptive procedure in Step 2 gives $\widehat\Sigma_u^{\mathcal{T}}$ such that the spectral error $\|\widehat\Sigma_u^{\mathcal{T}} - \Sigma_u\|  = O_P( m_p w_n^{1-q} )$ \citep{FanLiaMin11, CaiLiu11, RotLevZhu09} where we define $w_n = \sqrt{\log p/n}$. Furthermore, $\|(\widehat\Sigma_u^{\mathcal{T}})^{-1} - \Sigma_u^{-1}\| \le \|(\widehat\Sigma_u^{\mathcal{T}})^{-1} \| \|\widehat\Sigma_u^{\mathcal{T}} - \Sigma_u\| \|\Sigma_u^{-1}\|$. So $\|(\widehat\Sigma_u^{\mathcal{T}})^{-1} - {\Sigma_u}^{-1}\|$ is also $O_P( m_p w_n^{1-q} )$ due to the lower boundedness of $\|\Sigma_u\|$. So (\ref{eq3.4}) is valid.

Proving (\ref{eq3.5}) is trivial. $\|\widehat\Sigma_u^{\mathcal{T}} - \Sigma_u\|_{\infty} \le \|\widehat\Sigma_u^{\mathcal{T}} - \widehat\Sigma_u\|_{\infty} + \|\widehat\Sigma_u - \Sigma_u\|_{\infty} = O_P(\tau + w_n) = O_P(w_n)$ when $\tau$ is chosen as the same order $w_n$ and thus
\[
\|\widehat\Sigma^{\mathcal{T}} - \Sigma\|_{\infty} \le \|\widehat\Sigma_{12}\widehat\Sigma_{22}^{-1}\widehat\Sigma_{21}^T - B\Sigma_f B^T\|_{\infty} + \|\widehat\Sigma_u^{\mathcal{T}} - \Sigma_u\|_{\infty} = O_P(w_n)\,.
\]

Next let us take a look at the relative Frobenius convergence (\ref{eq3.6}) for $\|\widehat\Sigma^{\mathcal{T}} - \Sigma\|_{\Sigma}$.
\begin{equation} \label{eqA.2}
\begin{aligned}
\|\widehat\Sigma^{\mathcal{T}} - \Sigma\|_{\Sigma} \le & \;\|\widehat\Sigma_{12}\widehat\Sigma_{22}^{-1}\widehat\Sigma_{21}^T - \Sigma_{12}\Sigma_{22}^{-1}\Sigma_{21}^T\|_{\Sigma} +  \|\widehat\Sigma_{u}^{\mathcal{T}} - \Sigma_u\|_{\Sigma} \\
\le & \;\|(\widehat\Sigma_{12} - \Sigma_{12} )\widehat\Sigma_{22}^{-1}(\widehat\Sigma_{21} - \Sigma_{21})^T\|_{\Sigma} + 2\|(\widehat\Sigma_{12} - \Sigma_{12} )\widehat\Sigma_{22}^{-1}\Sigma_{21}^T\|_{\Sigma}  \\
& \;+ \|\Sigma_{12} (\widehat\Sigma_{22}^{-1} - \Sigma_{22}^{-1})\Sigma_{21}^T\|_{\Sigma} + \|\widehat\Sigma_{u}^{\mathcal{T}} - \Sigma_u\|_{\Sigma} \\
= &:\;\Delta_1 + 2\Delta_2 + \Delta_3 + \Delta_4\,.
\end{aligned}
\end{equation}
We bound the four terms one by one. The last term is the easiest,
\[
\Delta_4 \le p^{-1/2} \|\Sigma_{u}^{\mathcal{T}} - \Sigma_u\|_F \|\Sigma^{-1}\| = O_P( \|\Sigma_{u}^{\mathcal{T}} - \Sigma_u\|) = O_P(m_p w_n^{1-q})\,.
\]
Bound for $\Delta_1$ uses the fact that $\|\widehat\Sigma_{22}^{-1}\|$ and $\|\Sigma^{-1}\|$ are $O_P(1)$ and $\|\widehat\Sigma_{12} - \Sigma_{12} \|_F = O_P(\sqrt{p\log p/n})$. So
\[
\Delta_1 \le p^{-1/2} \|\widehat\Sigma_{12} - \Sigma_{12}\|_F^2 \|\widehat\Sigma_{22}^{-1}\| \|\Sigma^{-1}\| = O_P\Big(\frac{\sqrt{p}\log p}{n}\Big)\,;
\]
Bound for $\Delta_3$ needs additional conclusion that $\|\Sigma_{21}^T \Sigma^{-1} \Sigma_{12}\| \le \|B^T \Sigma^{-1} B\| \|\Sigma_{22}\|^2 \le 2\|\Sigma_{22}\| = O(1)$, where $B = \Sigma_{12} \Sigma_{22}^{-1}$ and the last inequality is shown in \cite{FanFanLv08}. So
\begin{align*}
\Delta_3 & = p^{-1/2} \tr^{1/2} \Big( (\widehat\Sigma_{22}^{-1} - \Sigma_{22}^{-1})\Sigma_{21}^T \Sigma^{-1} \Sigma_{12} (\widehat\Sigma_{22}^{-1} - \Sigma_{22}^{-1})\Sigma_{21}^T \Sigma^{-1} \Sigma_{12} \Big) \\
& \le p^{-1/2} \| (\widehat\Sigma_{22}^{-1} - \Sigma_{22}^{-1})\Sigma_{21}^T \Sigma^{-1} \Sigma_{12}\|_F \le p^{-1/2} \|\widehat\Sigma_{22}^{-1} - \Sigma_{22}^{-1} \|_F \|\Sigma_{21}^T \Sigma^{-1} \Sigma_{12}\| \\
& = O_P(\sqrt{\log p/(np)})\,.
\end{align*}
Lastly, by similar trick, we have
\begin{align*}
\Delta_2  &= p^{-1/2} \tr ^{1/2} \Big((\widehat\Sigma_{12} - \Sigma_{12} )\widehat\Sigma_{22}^{-1}\Sigma_{21}^T \Sigma^{-1} \Sigma_{21}\widehat\Sigma_{22}^{-1}(\widehat\Sigma_{12} - \Sigma_{12} )\Sigma^{-1}  \Big) \\
& \le p^{-1/2} \|\widehat\Sigma_{12} - \Sigma_{12} \|_F \|\widehat\Sigma_{22}^{-1}\| \|\Sigma^{-1}\|^{1/2} \|\Sigma_{21}^T \Sigma^{-1} \Sigma_{12}\|^{1/2} = O_P(\sqrt{\log p/n}).
\end{align*}
Combining results above, by (\ref{eqA.2}), we conclude that $\|\widehat\Sigma^{\mathcal{T}} - \Sigma\|_{\Sigma} = O_P(\sqrt{p}\log p/n + m_p(\log p/n)^{(1-q)/2})$.

Finally we show the rate of convergence for $\|(\widehat\Sigma^{\mathcal{T}})^{-1} - \Sigma^{-1} \|$. By Woodbury formula,
\[
\Sigma^{-1} = \Sigma_u^{-1} - \Sigma_u^{-1} \Sigma_{12} [\Sigma_{22} + \Sigma_{12}^T \Sigma_u^{-1} \Sigma_{21}]^{-1} \Sigma_{21}^T\Sigma_u^{-1} \,.
\]
Thus, let $A = \Sigma_{22} + \Sigma_{12}^T \Sigma_u^{-1} \Sigma_{21}, \widehat A = \widehat\Sigma_{22} + \widehat\Sigma_{12}^T (\widehat\Sigma_u^{\mathcal{T}})^{-1} \widehat\Sigma_{21}$ and $D = \Sigma_u^{-1} \Sigma_{12}, \widehat D = (\widehat\Sigma_u^{\mathcal{T}})^{-1} \widehat\Sigma_{12}$, we have the following bound similar to (\ref{eqA.2}):
\begin{equation} \label{eqA.3}
\begin{aligned}
\|(\widehat\Sigma^{\mathcal{T}})^{-1} - \Sigma^{-1} \| \le & \;\|\widehat D\widehat A^{-1}\widehat D^T - D A^{-1} D^T\| +  \|(\widehat\Sigma_{u}^{\mathcal{T}})^{-1} - \Sigma_u^{-1}\| \\
\le & \;\|(\widehat D - D )\widehat A^{-1}(\widehat D - D)^T\| + 2\|(\widehat D - D )\widehat A^{-1} D^T\|  \\
& \;+ \|D (\widehat A^{-1} - A^{-1}) D^T\| + \|(\widehat\Sigma_{u}^{\mathcal{T}})^{-1} - \Sigma_u^{-1}\| \\
= &:\;\widetilde\Delta_1 + 2\widetilde\Delta_2 +\widetilde \Delta_3 +\widetilde \Delta_4\,.
\end{aligned}
\end{equation}
From (\ref{eq3.4}), $\widetilde\Delta_4 = O_P(m_p \omega_n^{1-q})$. For the remaining terms, we need find the rates for $\|\widehat D - D\|$, $\|\widehat A^{-1}\|$, $\|D\|$ and $\|\widehat A^{-1} - A^{-1}\|$ separately. Note that $\|\Sigma_{12}\| = \|B \Sigma_{22}\| \le \|B\| \|\Sigma_{22}\| = O_P(\sqrt{p})$ by Assumption \ref{assmp::pervasive} (ii). So $\|D\| = O_P(\sqrt{p})$ and
\[
\|\widehat D - D\| \le \|(\widehat\Sigma_u^{\mathcal{T}})^{-1}\| \|\widehat\Sigma_{12} - \Sigma_{12}\| + \|\Sigma_{12}\| \| (\widehat\Sigma_u^{\mathcal{T}})^{-1} - \Sigma_u^{-1} \| = O_P(\sqrt{p} m_p \omega_n^{1-q})\,.
\]
In addition, it is not hard to show $\|\widehat A - A\| = O_P(p m_p \omega_n^{1-q})$. Also we claim $\|A^{-1}\| = O_P(p^{-1})$ since $\lambda_{\min}(A) \ge \lambda_{\min} (\Sigma_{12}^T \Sigma_u^{-1} \Sigma_{21}) \ge \lambda_{\min}(\Sigma_u^{-1}) \lambda_{\min}(\Sigma_f) \lambda_{r} (B\Sigma_f B^T)$ and by Weyl's inequality, $ \lambda_{r} (B\Sigma_f B^T) \ge  \lambda_{r} (\Sigma) - \|\Sigma\| \ge c p$ by Assumption \ref{assmp::pervasive} (i). Therefore, $\|\widehat A^{-1} - A^{-1}\| \le \|A^{-1}\| \|\widehat A^{-1}\|\|\widehat A - A\|$ implies $\|\widehat A^{-1} - A^{-1}\| = O_P(p^{-1} m_p \omega_n^{1-q})$, and furthermore $\|\widehat A^{-1}\| = O_P(p^{-1})$. Finally we incorporate the above rates together and conclude
\begin{align*}
\widetilde\Delta_1 &= O_P(p^{-1}\|\widehat D - D\|^2) = O_P(m_p^2 \omega_n^{2(1-q)})\,, \\
\widetilde\Delta_2 & = O_P(p^{-1/2}\|\widehat D - D\|) = O_P(m_p \omega_n^{1-q}) \,, \\
\widetilde\Delta_3 & = O_P(p\|\widehat A^{-1} - A^{-1}\|) = O_P(m_p \omega_n^{1-q})\,.
\end{align*}
So combining rates for $\widetilde\Delta_i, i = 1,2,3,4$, we show (\ref{eq3.7}) is true. The proof is now complete.

\end{proof}

\begin{proof}[\bf Proof of Theorem \ref{thm:asmp}]
Without loss of generality we can assume $\mu = 0$. By dominated converge theorem we know that for all $t$, $\lim_{n} \lambda_n(t) = -t$, that $\lambda_n(t)$ is differentiable, that $\lambda_n'(t) = - E \psi_n'(X-t)$, and that $\lim_n \lambda_n'(t) = -1$. With Taylor's expansion, we have
\begin{equation}\label{eqn:asmp:taylor}
\lambda_n(t) = \lambda_n(0) + \lambda_n'(0) t + \Delta_n(t),
\end{equation}
where $|  \Delta_n(t) | \le |t| \sup \{ | \lambda_n'(s) - \lambda_n'(0)| : 0 \le s \le t \}$.
Observe that
\begin{align*}
\big| \lambda_n'(s) - \lambda_n'(0) \big| &= \big| P(|X-s| \le \alpha_n) - P(|X| \le  \alpha_n) \big| \\
&\le P(|X-s| >  \alpha_n) + P(|X| >  \alpha_n).
\end{align*}
By Markov's inequality,
\begin{equation*}
\sup \{ | \lambda_n'(s) - \lambda_n'(0)| : 0 \le s \le t \} \le \frac{1}{\alpha_n}\big( 2 E|X| +| t | ).
\end{equation*}
For any $\epsilon \in (0,1)$, there exists $N > 0$, such that for all $n > N$,
\begin{equation*}
| \lambda_n(0)| \le 2, \quad \frac{1 + \epsilon/2}{1 + \epsilon} \le - \lambda_n'(0) \le \frac{1 - \epsilon/2}{1 - \epsilon}, \quad \frac{1}{\alpha_n}(2E|X|+4) \le \frac{\epsilon}{4(1+\epsilon)}.
\end{equation*}
Plugging $t = (1+\epsilon) \lambda_n(0)$ into (\ref{eqn:asmp:taylor}),
\begin{equation*}
\lambda_n((1+\epsilon) \lambda_n(0)) = \lambda_n(0) + \lambda_n'(0) (1+\epsilon)\lambda_n(0) + \Delta_n(t),
\end{equation*}
where $|\Delta_n(t)| \le (1+\epsilon) | \lambda_n(0)| \frac{\epsilon}{4(1+\epsilon)} = \epsilon| \lambda_n(0)| /4$. Equivalently,
\begin{equation*}
\lambda_n((1+\epsilon) \lambda_n(0)) = \lambda_n(0)( 1 + \lambda'_n(0)(1+\epsilon) + \beta_n),
\end{equation*}
where $|\beta_n| \le \epsilon / 4$.
Similarly,
\begin{equation*}
\lambda_n((1-\epsilon) \lambda_n(0)) = \lambda_n(0)( 1 + \lambda'_n(0)(1-\epsilon) + \beta'_n),
\end{equation*}
where $|\beta'_n| \le \epsilon / 4$. Also we have $1 + \lambda'_n(0)(1+\epsilon) + \beta_n < 0$ and $1 + \lambda'_n(0)(1-\epsilon) + \beta'_n > 0$. Multiplying both sides of the equations, we deduce that
\begin{equation*}
\lambda_n((1+\epsilon) \lambda_n(0)) \cdot \lambda_n((1-\epsilon) \lambda_n(0)) \le 0.
\end{equation*}
If $\lambda_n(0) = 0$, equation $\lambda_n(t) = 0$ has one zero $t=0$; and in fact it is the unique one for sufficiently large $n$, since $\lambda_n(t)$ is nonincreasing and $\lambda_n'(0) \neq 0$ for $n$ large enough. If $\lambda_n(0) \neq 0$, at least one zero lies in the interval with endpoints $(1+\epsilon) \lambda_n(0)$ and $(1-\epsilon) \lambda_n(0)$. Since $\lambda_n(0) \to 0$, for any zero $t'_n$ in this interval we have $t'_n \to 0$, which implies $\lambda'_n(t'_n) \to -1$. It follows that such zero is unique for sufficiently large $n$. This leads to $t_n / \lambda_n(0) \to 1$, thus proving the second claim in the theorem.

\medskip
The proof of the first claim is similar in spirit to that of \cite{Hub64}. Let us denote
\begin{align*}
T_n^{-} &= \sup\{ t: \sum_{i=1}^n \psi_n(x_i - t) > 0 \}, \\
T_n^{+} &= \inf\{ t: \sum_{i=1}^n \psi_n(x_i - t) < 0 \}.
\end{align*}
By monotonicity, $T_n \in [T_n^-, T_n^+]$. Since
\begin{equation*}
P(T_n^- < t) = P\big( \sum_{i=1}^n \psi_n(x_i - t) \le 0 \big),
\end{equation*}
it follows that for any fixed $z \in \mathbf{R}$,
\begin{align*}
P ( \sqrt{n}\, (T_n^- - t_n ) < z) &= P( T_n^- < t_n + z/ \sqrt{n} ) \\
&= P( \sum_{i=1}^n \psi_n (x_i - u_n ) \le 0 ) \\
& = P \Big( \frac{1}{\sqrt{n}} \sum_{i=1}^n \frac{ \psi_n(x_i - u_n) - \lambda_n(u_n)}{\sigma_n(u_n)} \le - \frac{\sqrt{n}\, \lambda_n(u_n)}{\sigma_n(u_n)} \Big),
\end{align*}
where we denote $u_n = t_n + z/\sqrt{n}$ and $\sigma_n(u) = E \psi_n(X-u)^2 - \lambda_n(u)^2$.

By dominate convergence theorem, $\lambda_n'(t_n) \to -1$ and $\sigma_n(u_n)^2 \to \sigma^2$. By Taylor expansion of $\lambda_n(u_n)$ at $t_n$,
\begin{equation*}
\lambda_n(u_n) = \lambda_n(t_n) + z/\sqrt{n} \, \lambda'_n(t_n) + \Delta^n_z,
\end{equation*}
where $|\Delta^n_z| \le n^{-1/2} |z| \sup\{ \lambda_n'(t_n + s) - \lambda_n'(t_n)|: 0 \le s \le z/\sqrt{n}\, \}$. A similar argument shows that
\begin{equation*}
\sup\{ \lambda_n'(t_n + s) - \lambda_n'(t_n)|: 0 \le s \le z/\sqrt{n}\, \} \le \frac{1}{\alpha_n}(2E(X) + 2|t_n| + |z|/\sqrt{n}\, ) = o(1).
\end{equation*}
This leads to $\lambda_n(u_n) = z/\sqrt{n}\, (\lambda_n'(t_n) + o(1)) = z/\sqrt{n}\, ( -1 + o(1))$, and thus $\sqrt{n}\, \lambda_n(u_n) \to -z$.

Let us write
$$\xi_i =  \frac{ \psi_n(x_i - u_n) - \lambda_n(u_n)}{\sigma_n(u_n)}$$
for the centered variance $\xi_i$ with unit variance. If we can show
\begin{equation}\label{eqn:asmp:normal}
\frac{1}{\sqrt{n}} \sum_{i=1}^n \xi_i \xrightarrow{d} N(0,1),
\end{equation}
then by continuity of $\Phi$, standard normal distribution function, we have
\begin{equation*}
P \Big( \frac{1}{\sqrt{n}} \sum_{i=1}^n \xi_i \le -\frac{\sqrt{n}\, \lambda_n(u_n)}{\sigma_n(u_n)} \Big) \to \Phi \big(\frac{z}{\sigma} \big),
\end{equation*}
which gives $P(\sqrt{n}\, (T_n^- - t_n) < z ) \to \Phi(z/\sigma)$. It is similar to show that $P(\sqrt{n}\, (T_n^+ - t_n) < z ) \to \Phi(z/\sigma)$. At this point, we are able to conclude that the first claim in the theorem holds, i.e.\ $\sqrt{n}\,(T_n - t_n) \xrightarrow{d} N(0, \sigma^2)$.

\medskip
To prove (\ref{eqn:asmp:normal}), it suffices to check Lindeberg's condition:
\begin{equation*}
E(\xi_i^2 \mathbf{1}\{ | \xi_i| > \sqrt{n}\, \epsilon\} ) \to 0
\end{equation*}
for any $\epsilon > 0$. Notice that $\lambda_n(u_n) \to 0$ and $\sigma_n(u_n) \to \sigma$, we only need to show
\begin{equation*}
E( \psi_n^2(X - u_n) \mathbf{1}\{ | \psi_n(X - u_n) | > \sqrt{n}\, \epsilon\} ) \to 0.
\end{equation*}
This is true due to
\begin{equation*}
\psi_n^2(X - u_n)  \le |X - u_n |^2 \le 2|X|^2 + 2u_n^2
\end{equation*}
and dominated convergence theorem.
\end{proof}

\bibliographystyle{ims}
\bibliography{Reference}

\end{document}